\documentclass[prb,aps,twocolumn,superscriptaddress,showpacs]{revtex4}

\usepackage{graphicx}

\begin{document}

\title{Spin dynamics near the critical doping in weakly-superconducting underdoped YBa$_{2}$Cu$_{3}$O$_{6.35}$ (T$_{c}$=18K)}

\author{C. Stock}
\affiliation{Department of Physics and Astronomy, Johns Hopkins University, Baltimore, Maryland 21218}

\author{W. J. L. Buyers}
\affiliation{National Research Council, Chalk River, Ontario, Canada K0J 1JO}
\affiliation{Canadian Institute of Advanced Research, Toronto, Ontario, Canada M5G 1Z8}

\author{Z. Yamani}
\affiliation{National Research Council, Chalk River, Ontario, Canada K0J 1JO}

\author{Z. Tun}
\affiliation{National Research Council, Chalk River, Ontario, Canada K0J 1JO}

\author{R. J. Birgeneau}
\affiliation{Department of Physics, University of California, Berkeley, CA 94720}
\affiliation{Canadian Institute of Advanced Research, Toronto, Ontario, Canada M5G 1Z8}

\author{R. Liang}
\affiliation{Physics Department, University of British Columbia, Vancouver, B. C., Canada V6T 2E7}
\affiliation{Canadian Institute of Advanced Research, Toronto, Ontario, Canada M5G 1Z8}

\author{D. Bonn}
\affiliation{Physics Department, University of British Columbia, Vancouver, B. C., Canada V6T 2E7}
\affiliation{Canadian Institute of Advanced Research, Toronto, Ontario, Canada M5G 1Z8}

\author{W. N. Hardy}
\affiliation{Physics Department, University of British Columbia, Vancouver, B. C., Canada V6T 2E7}
\affiliation{Canadian Institute of Advanced Research, Toronto, Ontario, Canada M5G 1Z8}

\date{\today}

\begin{abstract}

From neutron inelastic and elastic scattering we have determined the magnetic structure and fluctuations in the YBa$_{2}$Cu$_{3}$O$_{6.35}$ superconductor (T$_{c}$=18 K), near the boundary of the superconducting phase.  The long-range ordered collinear spins of the insulating antiferromagnet are replaced by a commensurate central mode arising from slow, isotropically polarized, short-range spin correlations extending over four planar unit cells. The inelastic spectrum up to 30 meV is also broad in wave vector and commensurate. In contrast to the the resonance peak of higher T$_{c}$ superconductors, the spins exhibit a single overdamped spectrum whose rate of relaxation $\Gamma$ decreases on cooling and saturates at $2\Gamma$=5 $\pm$ 1 meV below $\sim$ 50 K.  As the relaxation rate saturates the quasi-static spin correlations grow and become resolution limited in energy.  The spin susceptibility above $\sim$ 50 K follows the same $\omega/T$ scaling relation found for the monolayer La$_{2-x}$Sr$_{x}$CuO$_{4}$ system indicating that the dominant energy scale is set by the temperature. Below 50 K the scale length is geometric and not linked by velocity to dynamic widths. Despite the large differences from an antiferromagnet, we show that integrated intensity conserves the total moment sum rule, and that on cooling the spectral weight transfers from the inelastic spin relaxation to the elastic central peak.  There is no observable suppression of the spin fluctuations or central mode upon the onset of superconductivity. The spins respond not to coherent charge pairs but to hole doping allowing coexistence of glassy short range spin order with superconductivity.  Since the physics of the weakly superconducting system YBCO$_{6.35}$ must connect continuously with that in more strongly superconducting YBCO$_{6.5}$, we find that neither incommensurate stripe-like spin modulations nor a well-defined neutron spin resonance are essential for the onset with doping of pairing in a high temperature cuprate superconductor.

\end{abstract}


\maketitle

\section{Introduction}

There is a direct connection between high temperature superconductivity and antiferromagnetism.~\cite{Kastner98:70, Lee06:78, Kivelson03:75, Buyers06:386,Birgeneau:unpub}  Despite much effort, however, a universal theory behind the cuprate phase diagram has not been established.  One of the most interesting regions of the phase diagram lies near the antiferromagnetic phase and the onset of superconductivity.  This region with its low carrier concentration and highly suppressed transition temperature should provide much needed information on how antiferromagnetism is destroyed in favor of high-temperature superconductivity.

When the  superconducting order parameter is larger, it is known from an extensive study (Ref. \onlinecite{Stock04:69, Stock05:71}) in oxygen and structurally ordered  YBa$_{2}$Cu$_{3}$O$_{6.5}$ (YBCO$_{6.5}$ with T$_{c}$=59 K, corresponding to a hole doping of $p$=0.09), that the spin spectrum below and above T$_c$ extends to the lowest energies but with no elastic magnetic Bragg peak. The results do not support expectations that a new hidden order parameter might exist that would have created a spin gap and associated Bragg peak. The low-energy spin response is modulated incommensurately in momentum and rises to form an hour-glass shaped spectrum around a commensurate resonance at energy E=33 meV. For temperatures below T$_{c}$ the low-energy susceptibility is suppressed and its spectral weight is transferred to the resonance energy thereby conserving the total moment sum rule. Above T$_{c}$ the substantial broad response near resonance energies suggests incoherent pair formation.\cite{Stock04:69}  Experiments show that the resonance defines a crossover point that separates the low-energy incommensurate region from the high-energy region where the dynamics resemble spin-wave excitations.~\cite{Stock05:71,Hinkov04:430,Pailhes03:91}  Similar high-energy structure has been observed at larger doping in YBCO$_{6+x}$ (Hayden \textit{et al.}~\cite{Hayden04:429}) and in monolayer La$_{2-x}$Ba$_{x}$CuO$_{4}$ (Tranquada \textit{et al.}~\cite{Tranquada04:429}) and  La$_{2-x}$Sr$_{x}$CuO$_{4}$ (Christensen \textit{et al.}~\cite{Christensen04:93}).  Other studies of the spin excitation spectrum for similar hole doping in YBCO$_{6+x}$ by Fong \textit {et al.}~\cite{Fong00:61}, Dai \textit{et al.}~\cite{Dai01:63} and  Pailhes \textit{et al.}~\cite{Pailhes03:91} have mostly focussed on energies around the resonance.

Studies of La$_{2-x}$Sr$_{x}$CuO$_{4}$ (LSCO) for low carrier doping show that a spin-glass phase lies interposed between the antiferromagnetic insulating and the superconducting phases.~\cite{Matsuda00:62,Wakimoto00:62}  Such an intermediate phase has not been identified in YBCO. In LSCO the lightly doped antiferromagnetic insulator was found to break up into islands of correlated spins with moments pointing within the CuO$_{2}$ planes and the intensity from the ordered Cu$^{2+}$ moments decays rapidly on doping so as to enter the superconducting phase.~\cite{Wakimoto01:63,Fujita02:65}  This same phase and magnetic structure has recently been confirmed for overdoped La$_{2-x}$Sr$_{x}$CuO$_{4+y}$.~\cite{Mohottala06:5}  The superconducting and spin glass phases of LSCO both exhibit incommensurate peaks that are dynamic and static respectively.  Spin coupling to the lattice is more important for LSCO than for YBCO in view of the abrupt strongly first-order transition with doping from diagonal to collinear incommensurate wave vectors.~\cite{Fujita02:65}  The spin response follows $\omega/T$ scaling relation with temperature, $T$, over a broad range of energy $\omega$ with a breakdown observed at low-temperatures from the out-of-plane anisotropy.~\cite{Matsuda93:62} Inelastic scattering on Li doped La$_{2}$CuO$_{4}$~\cite{Bao03:91,Chen05:72}, where the Li causes a large perturbation as it replaces the Cu spin and charge in the superconducting CuO$_{2}$ plane, also shows a range of $\omega/T$ scaling.  Surprisingly it was suggested that the inelastic properties indicated the close proximity of a quantum critical point, in contrast to the spin-glass phase in LSCO and proximity to a known first order transition in the magnetic structure.

For YBCO there may be two classes of magnetic behavior as T$_{c}$ is suppressed. All studies for the fairly well doped YBCO$_{6+x}$ cuprates described above have shown that the magnetic spectrum differs considerably from that of the insulator. Its main features are the absence of elastic scattering, dynamic incommensurate fluctuations, a commensurate resonance, cones of highly damped spin waves and metallic resistivity.~\cite{Buyers06:386,Stock04:69,Stock05:71,Fong00:61,Dai01:63,Leyraud06:97,Lavrov98:41}  It is therefore of great importance to reduce further the hole doping to where the resistivity is insulating and then to observe how the antiferromagnetic phase evolves from the superconducting state and how the fluctuations are altered by the presence of mobile holes.  In particular, does the low-energy spin spectrum begin to resemble that of the antiferromagnet as the doping falls to the critical value p$_{c}$=0.055 where superconductivity disappears?  We will show that it does not in general. The similarities to the antiferromagnet are that the scattering becomes commensurate, that an elastic signal appears on cooling, and that the inelastic signal drops from a resonance region to very low energies. The differences are that the elastic and inelastic signals are broad in momentum signifying short-range correlations, that the elastic signal develops continuously without a transition, and that the spin directions remain isotropic.

Remarkably little neutron inelastic scattering has been done in YBa$_{2}$Cu$_{3}$O$_{6+x}$ (YBCO$_{6+x}$) for charge doping near the onset of superconductivity.   In YBCO$_{6.35}$ (T$_{c}$=39 K)  Mook \textit {et al.}~\cite{Mook02:88} observed incommensurate scattering at low-energies and a resonance at $\sim$ 23 meV.  Although the sample studied by Mook \textit{et al.} had the same nominal oxygen concentration as the crystal we have investigated, the in-plane doping was clearly much larger as shown by its large T$_{c}$=39 K.  Its properties place it in the fairly well-doped class described above. The sample we have studied has a more highly suppressed  superconducting transition temperature of T$_{c}$=18 K or $\sim$0.2 T$_{c}(optimal)=93 K$ and a lower toin-plane hole doping and thus lies much closer to the boundary of the superconducting phase.  It belongs in the low-doped class.

Most phase diagrams show that a cooled sample traverses a single boundary into the superconducting phase.  However, data from spin resonance of the positive muon~\cite{Sanna04:93,Niedermayer98:80} ($\mu$SR) in YBCO$_{6+x}$, in Ca doped YBCO and in LSCO, have been used to suggest a very different phase diagram.  Two additional  transitions are inferred within the superconducting dome, one to N\'{e}el order and at lower temperature a transition to a frozen glass.  We will show using neutron elastic and inelastic scattering that in YBCO$_{6.35}$ there is no transition to N\'{e}el order and that the nearly static Cu$^{2+}$ spins are isotropically polarized and have short correlation lengths.

Transport measurements have recently revealed some very interesting properties around the critical hole concentration for superconductivity.  Sutherland \textit{et al.} looked in YBCO and in LSCO for the electronic (linear-in-T) part of the thermal conductivity.~\cite{Sutherland05:94,Leyraud06:97}   When they reduced the doping to leave the superconducting phase they found YBCO to remain metallic whereas LSCO became insulating.  This marks a clear difference between the two systems over and above the spin coupling to rotation of the LSCO octahedra.  Liang \textit{et al.} observed a nonlinear relation between the lower critical field H$_{c1}$,  T$_{c}$ and the superfluid density in extremely underdoped  YBCO$_{6+x}$, in contrast to expectations from BCS theory.~\cite{Liang05:94} Photoemission results also show a direct correlation between electronic properties and hole doping, some of which have been correlated with the spin fluctuations.~\cite{Borisenko06:96}  Magnetoresistance measurements have found a large anisotropy in heavily underdoped YBCO$_{6+x}$ that has been attributed, without observational evidence, to the formation of charge stripes.~\cite{Ando99:83} It is therefore important to carry out a detailed neutron study in this region of low hole doping to determine if there is a Cu$^{2+}$ spin origin to the anomalous electronic properties.

\begin{figure}[t]
\includegraphics[width=80mm]{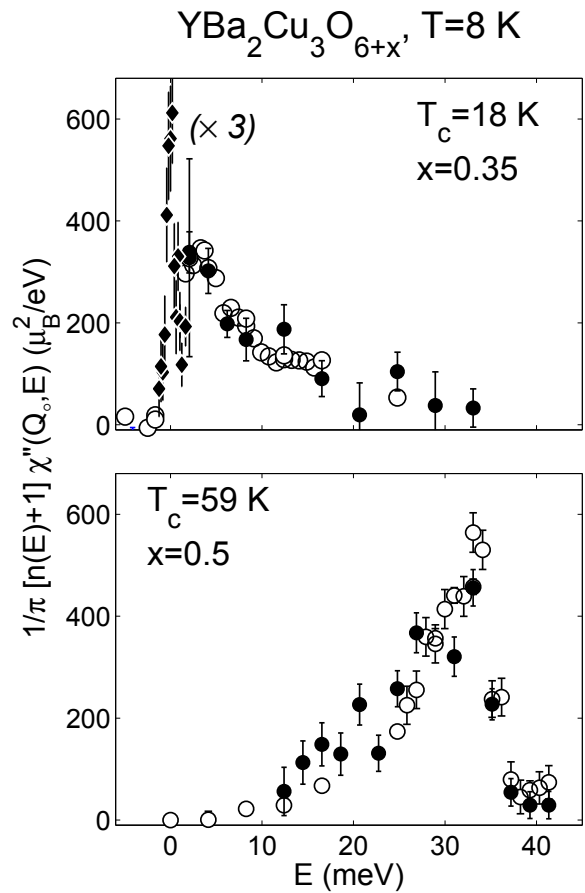}
\caption{The inelastic spectrum S({\bf{Q}},$\omega$) of YBCO$_{6.35}$ (T$_{c}$=18 K) compared with that of Ortho-II YBCO$_{6.5}$ (T$_{c}$=59 K) both measured in the superconducting state.  The momentum transfer is {\bf{Q}}=(0.5,0.5).  The spectral weight in YBCO$_{6.35}$ is concentrated at much lower energies relative to YBCO$_{6.5}$, but its integral is approximately conserved. Where no error in intensity is visible it is smaller than the symbol size because it is determined accurately from scans at constant energy. The closed (open) symbols are polarized (unpolarized) absolute intensities and the diamonds show the central mode intensity measured with polarized neutrons reduced by 3.}  \label{figure1}
\end{figure}

A general picture of how the spin structure and dynamics evolves as carrier concentration is reduced is given in Fig. \ref{figure1} which illustrates the qualitative change in the neutron scattering between Ortho-II YBCO$_{6.5}$ (T$_{c}$=59 K) and YBCO$_{6.35}$ (T$_{c}$=18 K) well below their superconducting temperatures. The spectral weight of  the intensity, proportional to S({\bf{Q}}, $\omega$) undergoes a dramatic change as doping is reduced. In YBCO$_{6.5}$ the response contains no elastic peak, and consists solely of a  continuum terminating at a well-defined resonance energy of $\sim$ 33 meV, while the response in YBCO$_{6.35}$ has been replaced  by an intense elastic peak and, at very low energy, a weak inelastic feature at $\sim$ 2.5 meV from spins that are relaxing not resonating. We will show a direct connection between the two, in which the relaxation rate of the inelastic feature slows on cooling and drives up the intensity of the central elastic component.  There is no detectable incommensurability.   As we shall see later, the total spectral weight, integrated over momentum and energy, is the same within error for the same energy range as that in the Ortho-II YBCO$_{6.5}$ superconductor, and therefore is the same as that of the antiferromagnet as shown earlier.~\cite{Stock05:71}  Thus the spin spectral weight is conserved as carriers are reduced towards the critical doping, despite the dramatic evolution of the spectral form.

This paper provides an overview in energy and temperature of the spin response measured with thermal and polarized neutrons, and it provides the foundation on which part of a previous short paper was based.~\cite{Stock06:73} There we focused on the scattering only around $\hbar \omega$=0.  Here we fully describe the relationship between the inelastic spin relaxation and the central peak as a function of temperature.  The first section gives the method, the second a detailed analysis of the central component, and the third section describes the higher energy spectrum up to energy transfers of 35 meV.  In the final section we discuss the integrated intensity and shows a direct connection between the scattering centered around $\hbar \omega$=0 and the broad inelastic scattering extending up to 30 meV at room temperature.

\section{Experiment}

Polarized and unpolarized neutron scattering measurements were made with the C5 DUALSPEC spectrometer at the NRU reactor, Chalk River Laboratories.  For unpolarized measurements a vertically focussing graphite (002) monochromator and a graphite (002) analyzer selected incident and scattered energies.  A pyrolytic graphite filter in the scattered beam eliminated higher order reflections, and a liquid nitrogen cooled sapphire filter before the monochromator reduced the fast neutron background.  Elastic measurements were made with 2 graphite filters after the sample while inelastic measurements were made with 1 filter.  Where we have compared elastic and inelastic intensities, we have corrected for the extra measured attenuation of the second filter.  For unpolarized measurements with fixed E$_{f}$=14.6 meV the horizontal collimation was [33$'$ 29$'$ \textit{S} 51$'$ 120$'$].  For unpolarized measurements with E$_{f}$=5 meV the collimation was [33$'$ 48$'$ \textit{S} 51$'$ 120$'$] and a liquid nitrogen cooled beryllium filter placed before the monochromator eliminated higher order neutrons. We have corrected all intensity the data for contamination of the incident beam monitor by higher wavelength neutrons as described elsewhere.~\cite{Stock04:69,Shirane_book}

To determine the magnetic structure and to extract the magnetic scattering over a broad range in energy transfer, we relied on polarized neutrons.~\cite{Moon69:181}  The  neutron beams were polarized with Heusler (111) crystals as monochromator and analyzer.  A graphite filter suppressed higher-order scattered neutrons.  A Mezei flipper in the incident beam allowed spin-flip (SF) and non-spin-flip (NSF) cross sections to be measured.  Two pairs of coils applied to the sample a weak 3-5 Gauss field to control the neutron spin direction, either a horizontal field (HF) parallel to ${\bf{Q}}$, or a vertical field (VF) perpendicular to ${\bf{Q}}$.  The flipping ratio was measured to be 12:1 for both configurations.  With the flipper on, the difference between the HF and VF count rates gave the spin-flip scattering from the magnetic electrons alone independent of nuclear incoherent or phonon scattering.  To change field and prevent any depolarization from trapped flux, the sample was always warmed to $\sim$ 50 K (more than twice T$_{c}$) and then field cooled to low temperatures in the new field.  Both field coils (and hence the field direction) were fixed with respect to the sample.

The sample consisted of seven $\sim$ 1 cc crystals, mutually aligned in the (HHL) scattering plane with an overall mosaic of $\sim$ 1.5$^{\circ}$.  From DC magnetization the critical temperature was measured to be T$_{c}$ = 18 K with a width of $\sim$ 2 K corresponding to a hole doping of $p$=0.06 (based on the formula provided by Tallon \textit{et al.}~\cite{Tallon95:51}).  This doping is close to that from a detailed study of  hole doping and lattice constants  by Liang \textit{et al.}~\cite{Liang06:73} The sample is orthorhombic with lattice constants $a$=3.843, $b$=3.871, and $c$=11.788 \AA\ and is twinned.  The same crystals, grown by a technique described previously~\cite{Peets02:15}, were used to obtain the low-energy properties with cold neutron  and the high-energy spectrum with spallation neutrons.~\cite{Stock06:73,Stock07:75}  Neutron diffraction revealed no well-defined oxygen superlattice Bragg peaks indicating that any oxygen ordering in the chains is short ranged.

To compare our data with those obtained from other oxygen concentrations and to determine the spectral weight, we normalized the intensity to an absolute scale by means of a constant-$Q$ scan through a transverse acoustic phonon near the (006) Bragg peak at (0.15,0.15,6) at 85 K.  This is the same procedure used previously where, by comparison to other calibration methods using a vanadium standard, we have found the systematic error in conversion to absolute units to be $\sim$ 15-20 \%.~\cite{Stock04:69,Stock05:71} To confirm the calibration via a phonon we have carried out a further independent calibration against the Bragg peak integrated intensity.  This procedure, used elsewhere~\cite{Shamoto93:48,Wakimoto01:63}, is described in the Appendix.  Having shown that the two methods agree well, we then quote the absolute intensities based on the phonon calibration.

\section{Elastic Scattering: Central peak}

\begin{figure}[t]
\includegraphics[width=80mm]{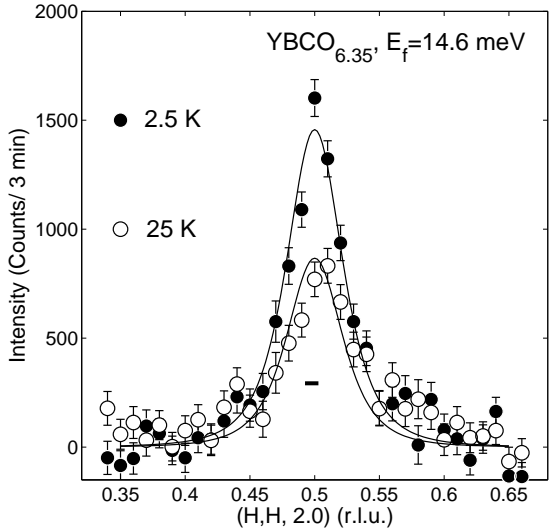}
\caption{In plane elastic scattering scans through the (1/2, 1/2, 2) magnetic peak at 2 K and 25 K with E$_{f}$=14.6 meV.  The magnetic intensity was obtained by subtracting a flat 80 K background scan.  The solid lines are results of fits to a Lorentzian squared as described in the text.  The horizontal bar represents the resolution width.  The inverse intrinsic width (and hence correlation length) does not diverge at low temperatures.} \label{elastic_4}
\end{figure}

We previously discovered that the inelastic spectrum at low temperatures exhibits two distinct energy scales, a slow central peak, energy-resolution limited as T$\rightarrow$0, and a fast timescale corresponding to broad overdamped inelastic fluctuations. We will show that the central peak exhibits three-dimensional (3D) correlations whereas the fast inelastic response displays only diffuse scattering from independent bilayers characteristic of two-dimensional (2D) correlations.  The onset of the central peak can then be seen as a crossover from 2D dynamic correlations to quasi-static 3D correlations.

\subsection{Unpolarized Experiments}

    With unpolarized neutrons we have determined the elastic magnetic scattering and in-plane correlations in scans of the form (H,H,2) and (H,H,5), and also determined the interplanar correlations with extensive scans along (1/2, 1/2, L). As background we subtracted the intensity remaining at 80 K as there is no significant temperature dependence in the quasielastic scattering above 80 K.  The results presented here have been confirmed with polarized neutrons (see later) which is sensitive only to the magnetic cross-section.  Typical scans through the correlated magnetic scattering are illustrated in Fig. \ref{elastic_4}.  The momentum broadening is much larger than resolution for the in-plane magnetic scattering at the (1/2, 1/2, 2) position and is similar at 2.5 and at 25 K. The strong elastic magnetic scattering observed in YBCO$_{6.35}$ contrasts with its absence in more highly doped Ortho-II YBCO$_{6.5}$.~\cite{Stock02:66}

\subsection{Correlation Lengths and Integrated Intensity}

    To extract quantitative information we fitted the data to a Lorentzian squared profile convoluted with the four-dimensional instrumental resolution function.  The following profile gave a good description of the in-plane q-dependence of the elastic commensurate magnetic scattering as illustrated in Fig. \ref{elastic_4}.

\begin{eqnarray}
\label{lor_sq}
S({\bf{Q,\omega}})=A\left({{1}\over{1+[{\bf{Q}-\bf{Q_{0}}}]^2}\times \xi^2}\right)^{2}\delta(\omega).
\end{eqnarray}

\noindent Here $A$ is an amplitude, ${\bf{Q}}=(2\pi/a)(H, K)$, ${\bf{Q_{0}}}=(2\pi/a)(0.5, 0.5)$, and $\xi$ is a correlation length that is isotropic within the $a-b$ plane.  As we shall see the out-of-plane correlations decay quite slowly along $L$ so that modeling the in-plane correlations with only the two-dimensional vector, ${\bf{Q}}$, is valid.  As the sample is twinned, we use tetragonal coordinates for the remainder of the paper.  Since a graphite analyzer was used only a limited integration over energy was done. However, since the central mode part of the spectrum is much narrower in energy at the lowest temperatures and stronger it follows that $\xi$ is to be interpreted not as the instantaneous correlation length but as one for near-static fluctuations.  This profile differs from the Lorentzian of our previous cold neutron work.  In the present  analysis the energy resolution was coarser (E$_{f}$=14.6 meV with $\Delta E$=1.5 meV full width at half maximum) so that we could observe the scattering up to higher energies and temperatures where it is weaker.

The Lorentzian squared profile follows from mean field theory for subcritical fluctuations.~\cite{Keimer92:46} The finite low-temperature correlation length can arise when random fields dominate over thermal fluctuations as previously applied to the spin-glass region of LSCO and in two-dimensional magnets.~\cite{Birgeneau83:28}  Both a Lorentzian and a Lorentzian squared give a good description of the data but the correlation range extracted is naturally shorter for the Lorentzian squared.  We note that in a previous work we used a Lorentzian squared to fit the correlations along the c-axis and a Lorentzian to fit the correlations within the a-b plane.~\cite{Stock06:73} The Lorentzian squared has the advantage of having a finite two-dimensional integral over momentum unlike a single Lorentzian.

From fits to scans along the [110] direction, convoluted with the resolution function, we extract an in-plane correlation length $\xi_{ab}$= 13 $\pm$ 2 \AA\ for the slow quasielastic fluctuations at 2.5 K.  Fits to the data along the [001] direction gave $\xi_{c}$=8 $\pm$ 2 \AA\ using the coupled bilayer model developed previously, which also corresponds to a Lorentzian-squared profile.~\cite{Stock06:73}  The in-plane and out-of-plane correlation lengths are the same within the limits of the accuracy.  Moreover we find that the width in wave vector of the elastic scattering is largely independent of temperature up to at least $\sim{50}$ K an example of which is shown in Fig. \ref{elastic_4} for 25 K.  The correlation lengths do not diverge with decreasing temperature as in insulating La$_{2}$CuO$_{4}$.~\cite{Birgeneau99:59}  We have therefore the interesting result that the central mode spin correlations are nearly isotropic and temperature independent.  The correlation length has a geometric rather than thermal origin, for it matches the hole spacing for a two-dimensional system, $l=(a/\sqrt{p})$ \AA $\sim$ 15 \AA\, where $a$ is the lattice constant and $p$ $\sim$ 0.06 is the hole doping. The in-plane correlation length is shorter than the indirect estimate of the mean free path (average of $l\sim$ 38 $\AA$ for YBCO$_{6.35}$) from optical conductivity~\cite{Lee04:70} but comparable to our Lorentzian analysis.~\cite{Stock06:73} Our results are similar to lightly doped and non-superconducting LSCO in its spin glass regime for concentrations of $\sim$ 0.02-0.04 for which $1/\xi$ also saturates at low-temperatures and increases only at very high temperatures.~\cite{Keimer92:46}  The difference here is that the short length scales coexist within the superconducting phase, and there is no observable incommensurate modulation. Incommensurate modulation has been observed in LSCO and YBCO and has been explained both in terms of stripes and quasiparticle excitations.~\cite{Kivelson03:75,Norman00:61,Norman00:61,Si93:47,Liu95:75,Kao00:61,Oleg01:63,Bascones05:xx,Liu03:90}  Based on the half-width at half maximum of our elastic peaks, we place an upper limit of $\sim$ 0.02 (defined as the peak splitting along [110]) for any possible incommensurability.  We note that an incommensurate wave vector of order $\sim$ 0.035 rlu (along [110]), as seen in superconducting LSCO (Ref. \onlinecite{Fujita02:65}) for a similar hole doping would have been readily detected with the resolution of this experiment (see Fig. \ref{elastic_4}). Though the incommensurability in YBCO$_{6+x}$ may scale differently with T$_{c}$ than in the monolayer LSCO system, our data could be described in terms of four peaks displaced from ($\pi$, $\pi$)if broadened in momentum.

\begin{figure}[t]
\includegraphics[width=93mm]{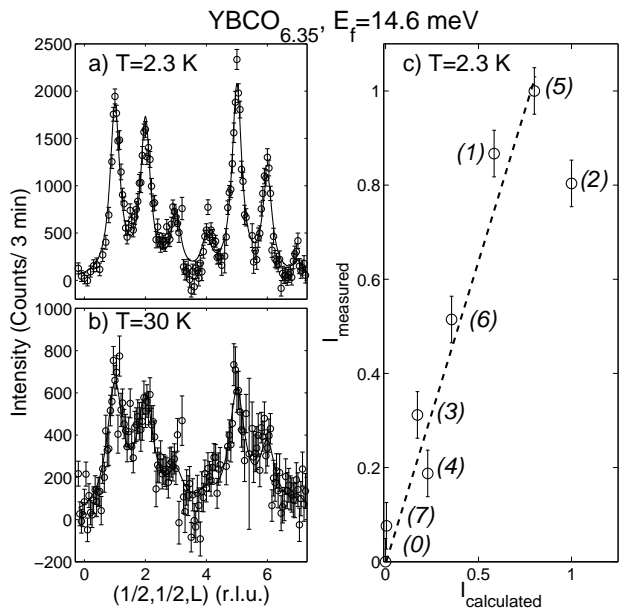}
\caption{$a)$ and $b)$ The elastic magnetic correlations along the [001] direction at 2.3 K and 30 K with a background at 80 K subtracted.  The correlation length along [001] is $\sim$ 1 unit cell.  The 3D correlations weaken above 30 K and arise from independent bilayers.  $c)$ The measured integrated intensity (defined in the text) is plotted as a function of the calculated integrated intensity for each labelled $L$ value.} \label{elastic_1}
\end{figure}

Scans along the [001] direction with E$_{f}$=14.6 meV are presented in Fig. \ref{elastic_1} at 2.3 and 30 K in panels $a)$ and $b)$.  A featureless background scan at 80 K has been subtracted.  The solid line is a fit to Lorentzians centered around each integer L values such that an intensity could be obtained for each L.  The measured integrated intensity at each L value is plotted as a function of the calculated intensity in panel $c)$.  The calculated intensity is based on the formula $I(L)=f^{2}(1/2,1/2,L)\sin^{2}(\pi \ L (1-2z))$ which is the product of the anisotropic Cu$^{2+}$ form factor and the bilayer structure factor with $z=0.36$.~\cite{Shamoto93:48} The measured integrated intensity at each integar L value was corrected for resolution effects as outlined elsewhere.~\cite{Cowley88:21}  Good agreement is obtained except for the L=2 peak which has too low an intensity.  While we do not currently understand why the L=2 peak is so weak, we note that on tightening the energy and momentum resolution considerably by going to E$_{f}$=5.0 meV, the ratio between the L=2 and L=1 peaks agrees perfectly with theory as does the ratio between L=2 and L=3.~\cite{Stock06:73}  We attribute this anomalously low intensity to a resolution effect. Even though peaks appear at the integer positions as expected for bilayers coupled ferromagnetically from cell to cell, the peaks are broad at all temperatures.  At high temperatures (30 K), the integer modulations weaken leaving two peaks centered on L=1.7 and 5.1, the fingerprint of the bilayer structure factor.  This indicates the effective absence of correlations between cells.  The c-axis correlation length of 8 $\AA$ may be thought of as a mean square transverse slippage by $\sim$ 2 $\AA$ of the in-plane spin correlations from cell to cell along c.

The diffraction pattern in Fig. \ref{elastic_1} leads to important conclusions. First it shows that the spins in the bilayer sheets are antiferromagnetically coupled at all temperatures, i.e., the phase across the bilayer is $\phi =\pi$.  Any sliding of the spin density in one layer relative to the other would be clearly visible.  Likewise any systemic phase difference between cells along c away from the in-phase $\phi =0$, would have caused diffraction peaks at non-integer L that are not seen.  Thus the regions of correlated spin density in one cell maintain a ferromagnetic in-phase relation from cell to cell.  The correlated spin region in one bilayer thus lies above the next cell along $c$. This is precisely the form of spatial fluctuation that precedes a critical quantum transition to 3D antiferromagnetism.  The correlation range is short and strongly sub-critical but the pattern is highly organized and not random.  If the spins were disordered in clusters as claimed ~\cite{Miller06:73,Sanna04:93} the regular integer-centered pattern would not be observed.  Neither disordered localized clusters, nor a cluster spin glass as inferred from muon studies, are in accord with experiment.  The spins clearly smoothly reorganize into a pattern where the susceptibility at 3D AF momenta ($L=n$) is a maximum.  The frozen 3D spin pattern we observe shows that a single phase exists in which superconductivity and glassy spins coexist.

At higher temperatures than those in Fig. \ref{elastic_1}, the 3D correlations become unobservable.  The elastic scattering becomes part of the single inelastic response discussed later in which the scattering is from independent bilayers.  Thus, both energy and temperature destroy the 3D correlations.

The saturation of the correlation length at low-temperatures is very similar to that observed in lightly doped LSCO.~\cite{Keimer92:46}  The very short correlations both within the plane and along [001], indicate that YBCO$_{6.35}$ is far from any critical point.  Our results also differ from the expected behavior near a quantum critical point where the correlation length should continuously diverge, not saturate at a high temperature value.  Even though further confirmation is needed at even lower oxygen doping, our results suggest that YBCO$_{6.35}$ is far from any quantum critical point. One possibility is that the transition out of the superconducting phase leads to a phase without long-range ordered antiferromagnetism. The possibility that the transition from  superconductor to a N\'{e}el ordered antiferromagnet may be first order is unlikely because of the smooth and continuous metallic thermal conductivity across the critical doping.~\cite{Sutherland05:94}

\subsection{Temperature Dependence}

\begin{figure}[t]
\includegraphics[width=90mm]{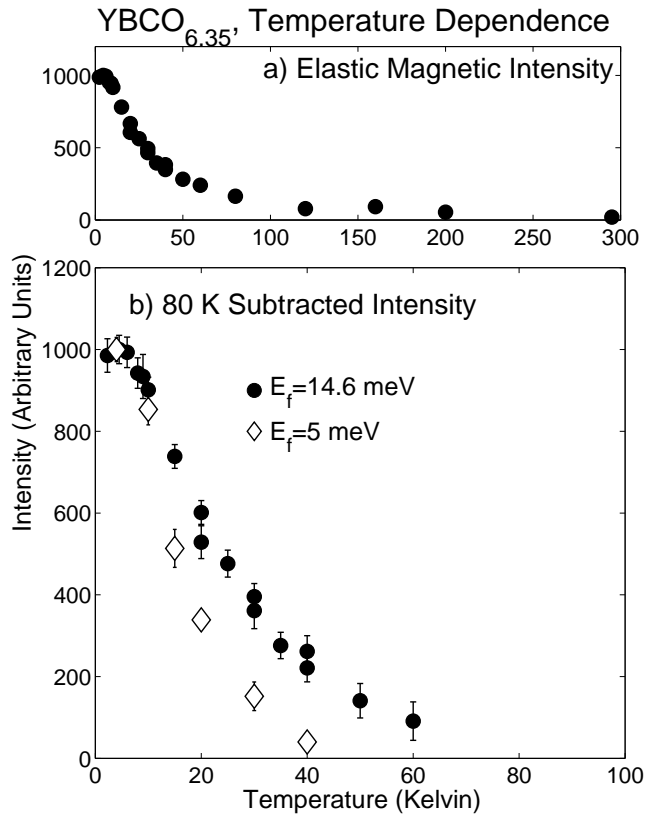}
\caption{$a)$ The temperature dependence of the peak magnetic intensity at $\hbar \omega$=0 with E$_{f}$=14.6 meV at $\bf{Q}$=(1/2,1/2,2.0).  $b)$ The temperature dependence of the elastic magnetic scattering measured with E$_{f}$=14.6 meV and 5 meV (C5 Chalk River) with a background measured at 80 K subtracted.  The wave-vector lies at the maximum of the broad spin correlations at $\bf{Q}$=(1/2, 1/2, 2.0) The different fixed final energies probe the magnetic fluctuations with different energy resolutions and hence different timescales.} \label{elastic_3}
\end{figure}

The temperature dependence of the elastic peak magnetic intensity was studied in momentum scans along the [110] direction through $\bf{Q}$=(1/2, 1/2, 2.0). The results would be similar for non-integer L as this is not a Bragg peak. As the elastic correlation length is known to be temperature independent (see previous section) the peak height of these scans represents the resolution integral over the spectral profile of the central mode at least as T$\rightarrow$0 where it is resolution limited.  The results of this analysis are presented in Fig. \ref{elastic_3}.  The directly measured magnetic intensity as a function of temperature at $\hbar \omega$=0 and at $\bf{Q}$=(1/2,1/2,2.0) is illustrated in panel $a)$.  The low temperature data was derived from scans along the [110] direction, examples of which are shown in Fig. \ref{elastic_4}.  High temperature data, where the central mode is weak, were extracted from inelastic constant-Q scans by extrapolating the magnetic intensity to $\hbar \omega$=0 as in Fig. \ref{mod_lor_1}.  Overlapping data taken at 80 K give good agreement between the two methods of extracting the elastic scattering in the absence of a strong central peak.  Since the elastic magnetic intensity shows little change above 80 K, in panel 4$b)$ we plot the magnetic intensity at $\bf{Q}$=(1/2,1/2,2.0) using 80 K as a background. For two different final energies, E$_{f}$=14.6 meV and 5 meV  with energy resolutions (full width at half maximum) of 1.5 meV and 0.1 meV respectively,  we find a decrease in the onset temperature on narrowing the energy resolution.  This demonstrates that very slow spin correlations occur as  in metallic spin-glasses which show a systematic lowering of the onset temperature as energy resolution is tightened.~\cite{Murani78:41}

The gradual onset of magnetic scattering in Fig. \ref{elastic_3}  contrasts with that observed in the insulator where a sharp well-defined Neel temperature is observed.~\cite{Shamoto93:48}  The growth of scattering with decreasing temperature might be thought to arise because more of the spin spectral weight, described by a single relaxation rate at each temperature, slows so as to fall within the energy resolution. This is an inconsistent view. If the central mode spectral profile were to correspond to a single Lorentzian relaxation rate, $\gamma(T)$, that slows on cooling, then its intensity would fall to half its zero temperature limit when $\gamma(T)$ matched the resolution half-width. For the 14.6 meV data this would indicate that $\gamma(T)= 0.75$ meV at T=20 K.  Evidence against a single Lorentzian relaxation rate, independent of the T-dependence of the static susceptibility, comes from cold neutron measurements~\cite{Stock06:73} where we know that the energy half width near 20 K is only 0.05 meV, not the much larger 0.75 meV width that would be required for a single relaxation rate model of the 14.6 meV data. As in a glass it is incorrect to associate either the half-height intensity or the apparent, sensitivity-dependent, high temperature onset with any single relaxation rate. The growth of central mode intensity seems to correspond to the appearance with cooling of a hierarchy of longer spin timescales.

Measurements at longer timescales and hence even finer energy resolution, $\sim5000$ times finer, have been made with $\mu$SR and have indicated a lower onset temperature of statically ordered moments.~\cite{Miller06:73,Sanna04:93} However the temperature where the peak has fallen to half its height has not decreased by such a large factor, but only to $\approx$10 K.  The much slower decrease in onset temperature with decreasing energy resolution indicates that the elastic correlations are neither statically ordered, nor fluctuating with a single relaxation rate that slows with decreasing temperature.

\subsection{Polarization Analysis}

We have verified the method of deriving the structure of the local moments from unpolarized data by measuring the polarized neutron cross-section since it is sensitive only to the magnetic cross-section.


\begin{figure}[t]
\includegraphics[width=80mm]{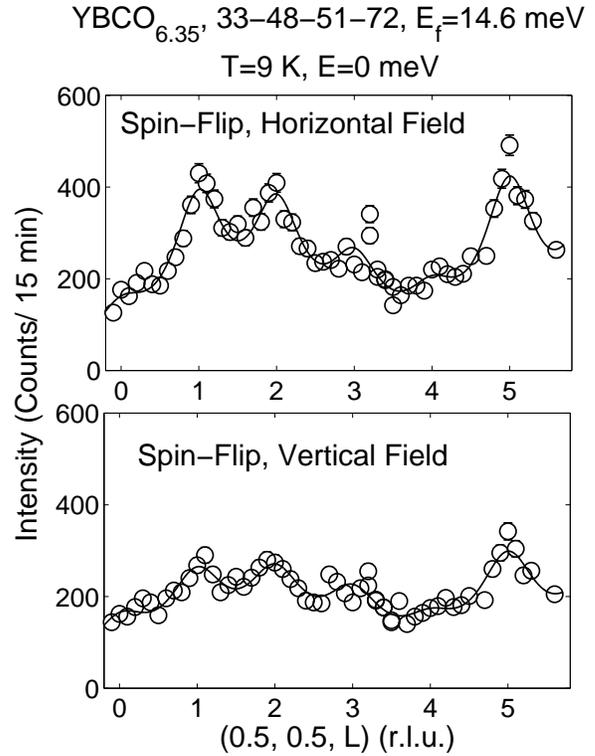}
\caption{Spin-flip cross-sections for horizontal field (HF) and vertical (VF) geometries measured at $\hbar \omega$=0.  For all values of L, the magnetic scattering in the HF channel is twice that in the VF channel, indicating no preferred orientation of the moments thus corresponding to a paramagnetic phase.} \label{polar_L}
\end{figure}

\begin{figure}[t]
\includegraphics[width=80mm]{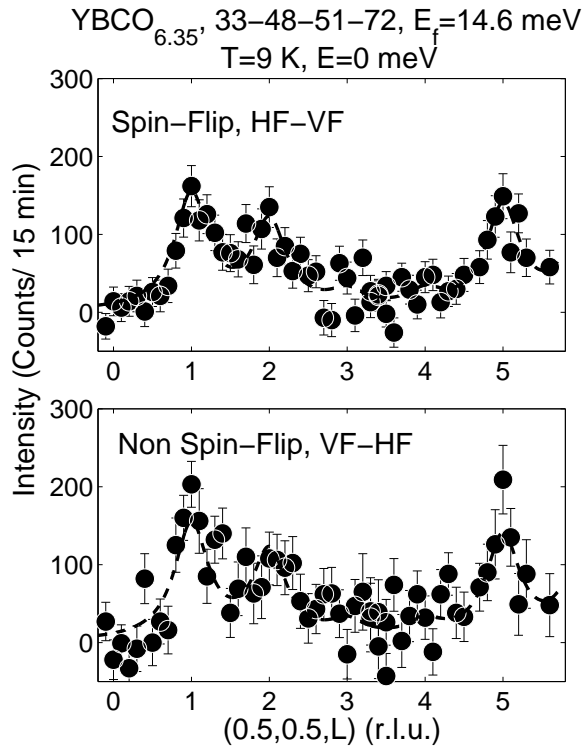}
\caption{Subtraction analysis of the polarized scans along [001].  The upper and lower panels both measure quasielastic magnetic intensity along the [1, $\overline{1}$, 0] or z spin direction but determined by two different methods.  The upper panel is from the spin-flip channel and gives $S_{zz}$, since for y perpendicular to z and to Q, HF=$S_{zz}+S_{yy}$ and VF=$S_{yy}$ for all $L$.  The lower panel is from the non spin-flip channel and, using the HF non-spin-flip intensity as background, it also gives $S_{zz}$, the fluctuations along the vertical $z$ direction.}
\label{polar_L_subtract}
\end{figure}

\begin{figure}[t]
\includegraphics[width=70mm]{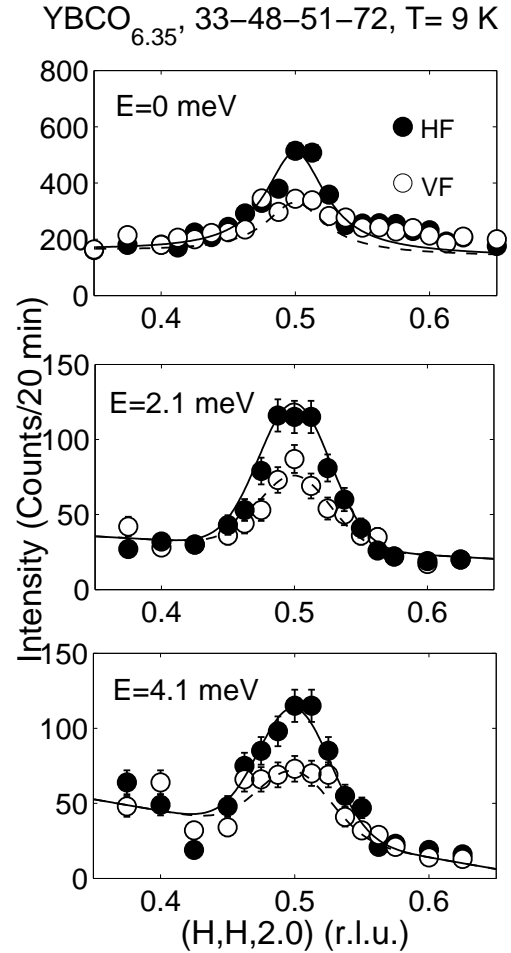}
\caption{Constant energy scans through the correlated magnetic peak in the superconducting state at 9 K.  The HF  cross-section (solid circles) is seen to be twice the VF cross-section (open circles) at each energy transfer.  The solid (broken) lines are Gaussian fits to the HF (VF) data that confirm that dynamic as well as static fluctuations are isotropic in spin direction.}
\label{polar_inelastic}
\end{figure}

\begin{figure}[t]
\includegraphics[width=90mm]{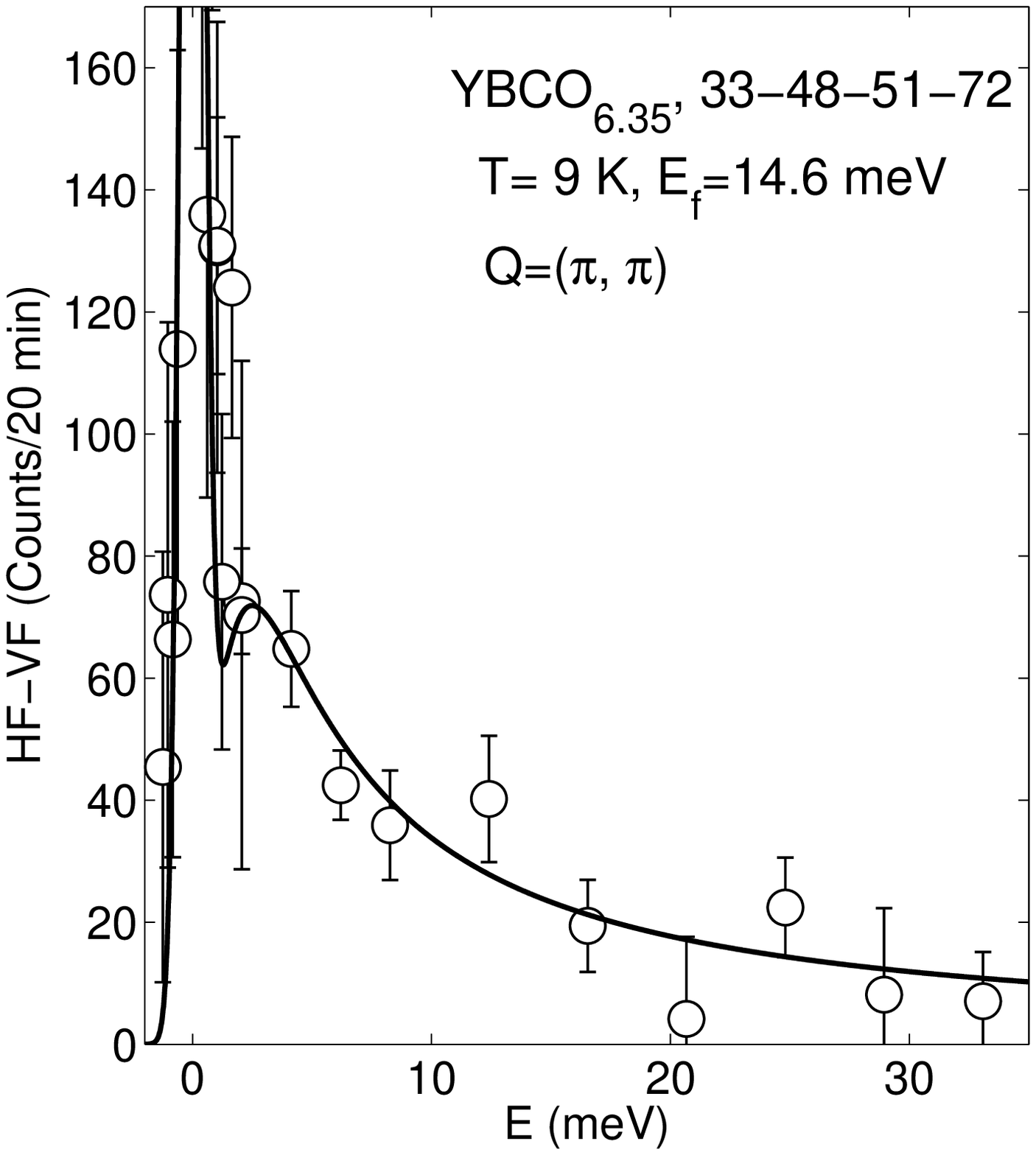}
\caption{The complete low-temperature inelastic and elastic spectrum obtained with polarized neutrons.  To obtain the purely magnetic scattering we have subtracted the VF from the HF as discussed in the text.  There are two spectral components, one centered around $\hbar \omega$=0 and another associated with the broad inelastic scattering extending to higher energies.} \label{spectrum_complete}
\end{figure}

In Fig. \ref{polar_L} the solid line shows a fit constrained so that the HF is twice the VF intensity.  The curve gives an accurate description of the data (the errors are about the size of the symbols).  Since the HF spin flip intensity is always twice the VF spin-flip intensity, it follows that there is no preferred orientation for the quasi-static moments.  The in-plane spin correlations along the vertical [1, $\overline{1}$, 0] $z$-direction were determined in two ways as shown in Fig. \ref{polar_L_subtract}. The upper panel shows the HF-VF intensity in the spin-flip channel, which gives the $S_{zz}$ spin scattering. It would amount to half of the total magnetic intensity for a paramagnet.  Because the nuclear contribution is normally large, the non-spin flip intensity is not usually used to determine magnetic cross-sections.  Nonetheless in the lower panel we show the subtraction VF-HF in the non-spin flip (NSF) channel.  The HF intensity contains no NSF magnetic scattering and is a measure of the (nuclear) background. The resulting non-spin-flip difference,  VF-HF,  measures only one of the two components, the same $S_{zz}$ of the magnetic moment perpendicular to {\bf{Q}}. It carries larger error because of the subtraction of two large nuclear signals.  The excellent agreement and consistency we obtain between spin-flip and non spin-flip methods adds credence to the analysis and experimental method.

Scans along the [110] direction through the correlated scattering are presented in Fig. \ref{polar_inelastic} for energy transfers of 0, 2.1, and 4.1 meV respectively.  The fitted lines are constrained such that the intensity in the HF channel is twice that in the VF channel.  The satisfactory description of experiment illustrates that the scattering cross-section is paramagnetic at all energy transfers.  There is no preferred direction for the spins, neither for the excitations nor for the static correlations.  The magnetic cross-section can then be extracted at all energy transfers by subtracting the VF from the HF scattering.  The paramagnetic cross-section at all energies is different than that observed in insulating YBCO where a significant out-of-plane anisotropy causes an energy gap of about 5 meV.~\cite{Tranquada89:40}

Our polarized neutron analysis of the elastic magnetic scattering  proves that the spin orientation (polarization) of the static moments is isotropic.  This is very different from the AF ordered insulator where the moments are pointing within the $a-b$ plane.  The spins in the superconductor form a paramagnetic structure that we call a hedgehog phase. The isotropic polarization of the spins is not surprising given that the correlation lengths are nearly isotropic with the in-plane close to the out-of-plane correlation length.  We then only expect significant anisotropy to develop when the correlation length grows to more than a few lattice constants in the $a-b$ plane and along the $c$ direction.  Our results may be consistent with theoretical predictions of Hasselmann \textit{et al.} (Ref. \onlinecite{Hasselmann04:69}) and Lindgard (Ref. \onlinecite{Lindgard05:95}) based on spiral spin states. They predict the lightly doped region of the cuprate phase diagram is characterized by a significant dipolar frustration of the antiferromagnetic order leading to spiral correlations.  The frustration is caused by holes (which reside primarily on the oxygen atom) causing Cu$^{2+}$ spins locally to be ferromagnetically coupled. Earlier strong ferromagnetic exchange through holes was found to cause chiral spin rotations  and destroy order.~\cite{Aharony88:60} Models for the dispersion by Lindgard~\cite{Lindgard05:95}  could result in an out of plane component to the moment as observed in YBCO$_{6.35}$.

Fig. \ref{spectrum_complete} shows the complete low-energy spectrum of YBCO$_{6.35}$ at 9 K using polarized neutrons.  The magnetic cross-section for energy transfers greater than 2 meV was extracted from constant energy scans along the [110] direction over the ridge of scattering.  The two channels were then subtracted and a Gaussian  fitted to the HF-VF data.  For for energies less than 2 meV a single point was measured at (1/2, 1/2, 2.0) in both the HF and VF channels and subtracted.   The spectrum clearly shows the spins fluctuate on two time scales, one giving a sharp central elastic peak  and the other a broader component peaked around 2 meV and smoothly decreasing with energy up to $\sim$ 35 meV (the highest energy measured).  The spectral profile agrees with that derived from unpolarized neutrons as shown in Fig.  \ref{figure1} and as discussed in the next section.

\section{Inelastic response}

In the previous section, we discussed the temperature and momentum dependence of the elastic and low energy magnetic scattering.  The lowest energy fluctuations within the central peak arise from spin correlations that, for the lowest temperature, become quasi-static on the neutron timescale.  While their correlation lengths within the basal a-b plane and along the c-axis are short, these slow fluctuations are 3D in nature. In this section, we discuss the inelastic scattering from faster spin fluctuations.  We will show that the fast fluctuations are primarily two-dimensional in nature and will describe the temperature dependence of their profile in energy and momentum.

\subsection{Momentum profile}

\begin{figure}[t]
\includegraphics[width=80mm]{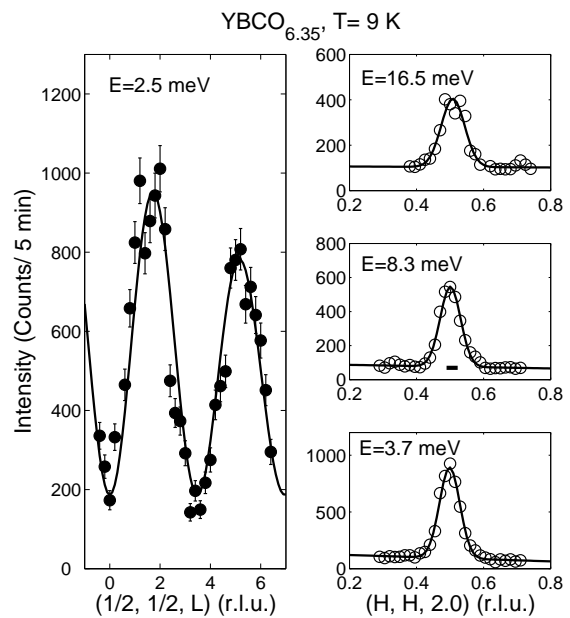}
\caption{Out-of-plane and in-plane dynamic magnetic correlations showing the momentum dependence of the magnetic scattering at T=9 K for various energy transfers.  The solid line for the scan along (1/2, 1/2, L) gives the behavior expected for independent bilayers.  The solid lines through the (H, H, 2) scans are results of fits to Gaussians.  The horizontal bar represents the momentum resolution along the [110] direction.}
\label{momentum}
\end{figure}

\begin{figure}[t]
\includegraphics[width=93mm]{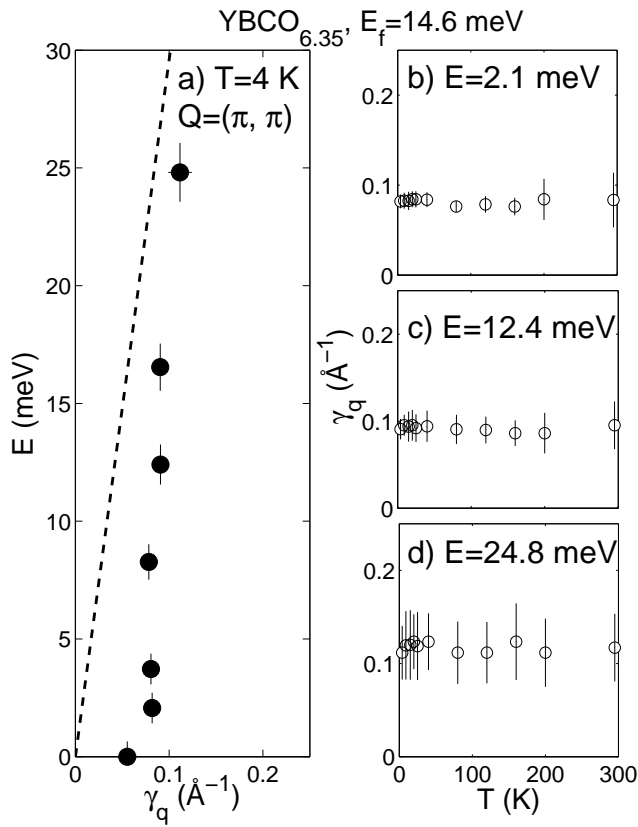}
\caption{$a)$ The half width at half maximum in momentum as a function energy transfer at 4 K. The dashed line in the left-hand panel illustrates the spin-wave velocity (hence E=$\hbar c \gamma_{q}$=$590 \gamma_{q} $) determined at high energy (Ref. \onlinecite{Stock07:75}).  $b)-d)$ The temperature dependence of the half width at half maximum at energy transfers of 2.1, 12.4, and 24.8 meV.  The widths are from fits of a Gaussian profile to scans along  (H,H) direction around ${\bf{Q}}$=(1/2,1/2).} \label{q_width}
\end{figure}

The momentum dependence of the correlated magnetic scattering is illustrated in Fig. \ref{momentum} with scans along the [001] and [110] directions.  The solid lines through the scans along the (H, H, 2) direction are results of fits to a Gaussian of fixed width convolved with the resolution function. The good agreement shows that the in-plane correlation length is largely independent of energy just as earlier we showed its independence of temperature. The solid line through the scan along the (1/2, 1/2, L) direction is the result of a fit to the following equation representing the acoustic bilayer structure factor:

\begin{eqnarray}
\label{eq1_mod_lor}
S(L)=A f^{2}(L) \sin^{2}(\pi \ L (1-2z))+B.
\end{eqnarray}

\noindent Here $A$ is an amplitude, $(1-2z)$ represents the bilayer spacing with $z$=0.36, $f^{2}(L)$ is the anisotropic Cu$^{2+}$ form factor, and B is a constant. This model with two free parameters gives an excellent account of the data over a wide range in L. The spin excitations therefore arise from independent bilayers.  Note that the optic fluctuations play no role as they lie above an onset of $\sim$ 50 meV (Ref. \onlinecite{Stock07:75}), well outside the energy range of this experiment.  It is interesting that the peaks which appeared at integral values of L for the quasielastic central peak are no longer observable.  Therefore, the 3D behavior is only exhibited by the slow dynamics of the central peak, whose large susceptibility enhances 3D correlations, and not by the faster timescale of the inelastic channel.

The observed in-plane magnetic scattering is commensurate. There is no sign of the flat-topped structure nor of the hour-glass dynamic incommensurability observed at larger doping~\cite{Stock05:71} in Ortho-II YBCO$_{6.5}$,  nor of the well-defined incommensurate features of lightly doped normal and superconducting LSCO~\cite{Fujita02:65} and LBCO.~\cite{Fujita04:70}  Instead Fig. \ref{momentum} shows a simple in-plane momentum broadening that is very weakly dependent on energy between 2.1-24.8 meV with an average Gaussian-fitted half width at half maximum (hwhm) (Fig. \ref{q_width}) of 0.09 $\pm$ 0.02 ${\AA}^{-1}$.  Indeed the scans at various energies in Fig. \ref{momentum} are well described by a constant width. The very small increase of momentum broadening with energy (0.0.09 at 3.7 meV to 0.10 ${\AA}^{-1}$ at 16.5 meV) mirrors the slope from the large spin-wave velocity which is known from high-energy measurements on the same sample~\cite{Stock07:75} and is indicated by the dashed line in Fig. \ref{q_width} $a)$. In contrast, for larger doping the momentum broadening decreases with energy because of the hour-glass dispersion.~\cite{Stock05:71} For E=0 the Gaussian half width (hwhh) is 0.06 $\pm$ 0.01 ${\AA}^{-1}$ or about two thirds of the broadening of the spin excitations.

Apart from momentum broadening, the commensurate profile along [110] resembles that in the insulating region of YBCO and LSCO  despite the fact that YBCO$_{6.35}$ is metallic and superconducting.  For systems with a highly suppressed T$_{c}$ such as YBCO$_{6.35}$ a dynamic stripe model may therefore not apply. It is possible that, for very low concentrations of holes, regions of correlated spins are favored over a quasi-one dimensional alternation of spins and charge.  We note, however, that the expected value of the incommensurability (denoted as $\delta$) is small.  Based on the monolayer LSCO system~\cite{Yamada98:57}, where the incommensurability for 6 $\%$ doping is $\sim$ 0.05 r.l.u. or 0.08 ${\AA}^{-1}$ along [1 0 0], we would expect for our sample an incommensurate wave vector of 0.06 ${\AA}^{-1}$ or 0.025 in H as our scan along [H H 0] passes between putative incommensurate peaks.  This is close to the hwhh of 0.06 $\pm$ 0.01 ${\AA}^{-1}$ or 0.025 in H we measure for E=0 correlations (Fig.\ref{elastic_4}). Since the elastic fwhm resolution in H is 0.01, an incommensurate wave vector of 0.025 would have been easily resolvable if it were well-defined. Thus the absence of incommensurable peaks stems not from the more than adequate resolution but from sources of intrinsic broadening such as the short range of the oxygen chain order or the dominance of magnetism over superconductivity when T$_{c}$ is suppressed.

\subsection{Energy spectrum}

\begin{figure}[t]
\includegraphics[width=80mm]{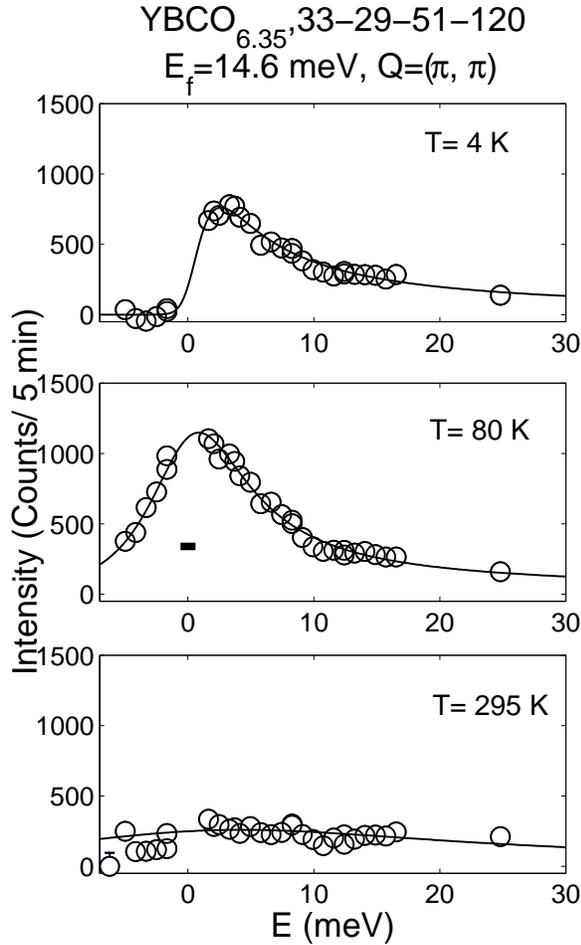}
\caption{The magnetic scattering as a function of energy transfer is plotted for 4, 80, and 295 K.  The data was obtained using unpolarized thermal neutrons with a non-magnetic background subtracted.  The solid line is the result of a fit to the modified Lorentzian lineshape described in the text.  The data around $\hbar \omega$ =0 has been removed so that the purely inelastic spectrum can be analyzed free from the central quasielastic component.  The horizontal bar is the energy resolution at the elastic line. The elastic scattering around $\hbar \omega$=0 has been removed.} \label{mod_lor_1}
\end{figure}

\begin{figure}[t]
\includegraphics[width=80mm]{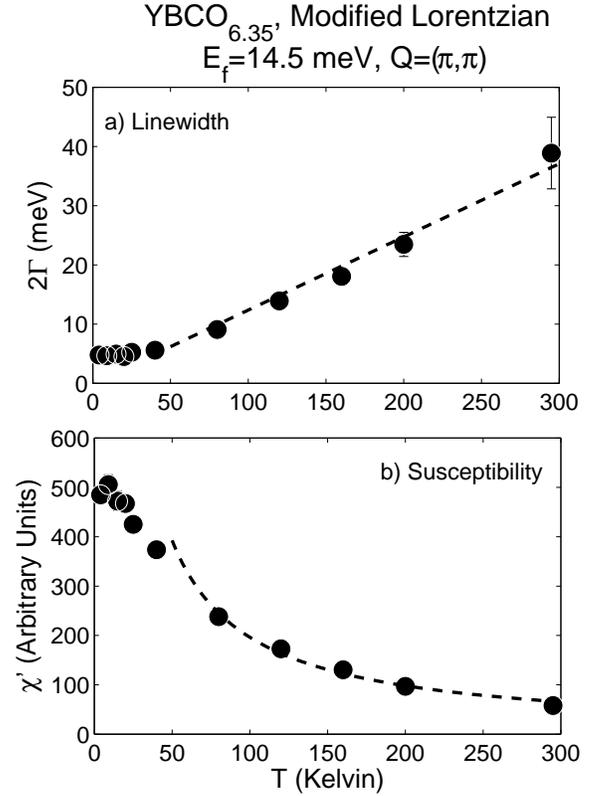}
\caption{The temperature dependence of the linewidth 2$\Gamma$ and the susceptibility are plotted.  The broken line in panel \textit{a)} is the result of a fit to a straight line through the origin.  The broken line in panel \textit{b)} is the result of a fit of the high temperature data to $1/T$.  This temperature behavior breaks down for temperature below $\sim$ 50 K.}
\label{mod_lor_2}
\end{figure}

We determined the energy dependence of the magnetic scattering from constant {\bf{Q}} and from  constant energy scans for energy transfers up to $\sim$ 25 meV.  The peak magnetic intensity was extracted from the constant {\bf{Q}} scans on the ridge at (H,H)=(0.5, 0.5) from which a background averaged between (0.3, 0.3) and (0.7, 0.7) was subtracted.  An independent determination of the magnetic signal was obtained at selected energies in scans over the ridge along [110] direction such as those in Fig. \ref{momentum}. The peak magnetic intensity from a Gaussian fit gave excellent agreement with the constant $\bf{Q}$ method.

The magnetic continuum spectrum thus determined  exhibits spin relaxation (Fig. \ref{mod_lor_1}) that broadens and weakens as temperature rises.  At low temperature the scattering is peaked at around 2 meV, the limiting relaxation rate for the spin excitations for the short-range or glassy spins.  The 2 meV relaxation rate lies much below any energy derived from the product of momentum width and high spin velocity. At 295 K the scattering is broad and almost energy independent.  The absence of data around $\sim$ 20 meV is caused by a strong phonon which is difficult to correct for with unpolarized neutrons.  We know, however, that the magnetic scattering continues smoothly through the phonon region from the polarized neutron analysis (Fig. \ref{spectrum_complete}).  The spectrum differs considerably from that observed in YBCO$_{6.5}$ (T$_{c}$=59 K) at low temperatures (Fig. \ref{figure1} which rises to a resonant peak in the susceptibility at 33 meV~\cite{Stock04:69}.  By suppressing the critical temperature from 59 K to 18 K, the spectral weight has shifted to low-energy at temperatures well below the onset of superconductivity. Instead of the positive slope of the dynamic susceptibility for the 59 K superconductor as the energy rises to a 33 meV maximum at a resonance, we find a negative slope corresponding to a weakening susceptibility as the energy rises beyond the maximum of 2 meV.  These two very different behaviors require a major electronic reorganization to have taken place.

To obtain the relaxation rate and amplitude of the spectrum at Q=(1/2,1/2) and below $\sim$ 25 meV we fitted the scattering, with the quasielastic central peak removed, to the following modified Lorentzian, which corresponds to overdamped excitations with a single energy scale, the relaxation rate $\Gamma$,

\begin{eqnarray}
\label{eq2_mod_lor}
S({\bf{Q}},\omega)=\chi'[n(\omega)+1]{{\omega\Gamma}\over{\Gamma^2+\omega^2}}.
\end{eqnarray}

\noindent where $\chi'$ is the static susceptibility, and $[n(\omega)+1]$ is the Bose factor..  The solid lines in Fig. \ref{mod_lor_1} shows that this model with only two adjustable parameters at each temperature gives an excellent description of experiment over a very broad range in temperature and energy. The fast spin dynamics therefore relax with a single temperature dependent time constant.  The amplitude and relaxation rate are shown in Fig.\ref{mod_lor_2} up to room temperature.

As illustrated in Fig. \ref{figure1} the spectrum of spin fluctuations at low temperatures in YBCO$_{6.35}$ (T$_{c}$=18 K) differs considerably from that observed in Ortho-II YBCO$_{6.5}$.  YBCO$_{6.5}$ displays a clear resonance peak at $\sim$ 33 meV in the superconducting phase.  High-energy measurements in YBCO$_{6.5}$ have shown the resonance defines a cross-over point between low-energy anisotropic incommensurate fluctuations and high-energy spin-wave-like fluctuations.~\cite{Stock05:71}  In contrast to this hour-glass shaped dispersion, in YBCO$_{6.35}$, no clear resonance peak is observed.  Even though the neutron scattering response peaks at $\sim$ 2 meV the spin response is relaxational, and  does not scale with T$_{c}$ as does the sharp resonance peak in YBCO$_{6.5}$.  The relaxational feature does carry, however, a substantial fraction of the low-energy spectral weight.

The absence of a spectrally sharp resonance can be attributed to several effects.  One possibility is that spin disorder introduced through short oxygen chains strongly damps the resonance peak.  We find this conclusion unlikely, for if the chain disorder were to introduce such a large effect as to completely destroy a resonance peak, we would expect a strong effect on T$_{c}$ (which is measured to be very well defined in our crystals).  We would also expect that, in a completely disordered sample, the resonance peak would be absent.  This is clearly not the case as a resonance peak is observed in oxygen disordered samples near the YBCO$_{6.5}$ concentration and also in heavily overdoped materials.~\cite{Fong00:61}   While it has been found that impurity doping affects the width of the resonance peak, the ratio of the resonance energy to that of T$_{c}$ remains unchanged with impurity doping.~\cite{Sidis00:84}  Scaling the resonance energy with T$_{c}$, from our YBCO$_{6.5}$ results~\cite{Stock04:69} we would expect a resonance peak at 10 meV in YBCO$_{6.35}$, clearly not the case in our measurements.  Therefore, chain disorder cannot completely account for the absence of a strong resonance peak in YBCO$_{6.35}$.  Spectral broadening is known  to occur, for example in YBCO$_{6.5}$, where disorder increases the resonance width at low temperatures from 10 meV in ordered Ortho-II to 25 meV in a disordered sample.~\cite{Fong00:61} It has been found that while impurity doping increases the width of the resonance peak, the ratio of the resonance energy to that of T$_{c}$ remains unchanged with impurity doping.~\cite{Sidis00:84}  The 2 meV feature may possibly be viewed in terms of depression by large damping near the critical doping as the spin response evolves into critical scattering.

A low resonance energy might also arise because the incommensurate wave vector is small or absent.  If we adopt the stripe model~\cite{Kruger04:70}, where the resonance is a cross-over energy that separates incommensurate fluctuations from spin-waves, we would then expect a very low energy resonance because of the low incommensurability.

Our observation of strong but only relaxational spin excitations in YBCO$_{6.35}$ indicates that any  resonance does not need to be well-defined for superconductivity to occur. However spectral weight from the resonance region can be important in heavily doped systems such as YBCO$_{6.95}$ where it is the only easily detectable magnetic spectral feature and where it gives more than enough change in magnetic energy to account for the superconducting condensation energy.~\cite{Woo06:2}  The belief that a resonance cannot account for superconducting pairing exists because the resonance comprises so little of the total spectral weight (Ref. \onlinecite{Kee02:88}). Other theories link the resonance to the pairing boson (Refs. \onlinecite{Abanov02:89}, \onlinecite{Eschrig03:67}).  The decoupling of the resonance energy from T$_{c}$ is illustrated by LBCO at a doping of 1/8 where the resonance energy occurs at $\sim$ 60 meV despite the fact that T$_{c}$ is almost zero.~\cite{Tranquada04:429}    Thus the resonance energy may scale with T$_{c}$ in YBCO over a limited range of hole doping, but it does not near 1/8 doping nor for different cuprates.~\cite{Tranquada04:429} In LSCO two peaks have been observed in $\chi''$ and the lower energy peak is not obviously connected with cross over from stripe-like fluctuations to spin waves.~\cite{Vignolle07:03}  Instead the resonance energy may scale with the incommensurability that is driven by accommodation of holes rather than by pairs, and hole doping is small for YBCO$_{6.35}$.

The susceptibility $\chi'$ and full-width 2$\Gamma$ as a function of temperature are plotted in Fig. \ref{mod_lor_2}.  Panel \textit{b)} shows that the susceptibility on cooling initially grows strongly as  $1/T$ (broken line), then slows below 50 K and finally saturates below $\sim$ 10 K. Panel \textit{a)} shows that the linewidth (full width at half maximum) decreases linearly with temperature until it saturates below $\sim$ 50 K.   The broken line in Panel \textit{a)} is a high temperature fit to

\begin{eqnarray}
\label{highT_scale}
2\Gamma=(1.40\pm0.15) k_{B}T,
\end{eqnarray}

\noindent where $\Gamma$ and $k_{B}T$ are given in meV.  The proportionality of the relaxation rate to temperature implies that temperature is the dominant energy scale above $\sim$ 50 K.  As shown in the next section, $\omega/T$ scaling is expected and found to be obeyed over a broad range in temperature and energy.

For critical scattering in a two dimensional periodic Heisenberg system, the energy linewidth ($\Gamma$) is related to the correlation length ($\xi$) by $\hbar \Gamma=\hbar c/\xi$ where $\hbar c$ is the spin-wave velocity.~\cite{Chou91:43,Tranquada90:64}  Saturation of the characteristic energy linewidth would then be attributed to saturation of the correlation length.  Instead, as discussed above, the spatial correlation length determined from constant-energy scans is almost independent of temperature below 50 K (Figs. \ref{elastic_4}, \ref{scaling_2}).   This indicates that the length is geometrically limited, but its magnitude is not linked by velocity to the dynamic width.  For example the 4 K correlation length of 1/$\xi$ = 0.09 ${\AA}^{-1}$ times the velocity of 590 meV-$\AA$ gives an energy of 50 meV, much too large to account for the 2 meV relaxation rate. Thus any relation between spatial and temporal scales breaks down for the spin relaxations and more so for the spectrally narrower central peak.  Instead measurement shows that  $\Gamma$ saturates at the same temperature where the spectral weight transfers to the much slower central peak, as discussed later,  with an almost constant correlation length for both.  A similar result was obtained in  La$_{2-x}$Sr$_{x}$CuO$_{4}$ where the correlation length saturates at low temperatures for a similar hole doping.~\cite{Keimer91:67}  Any effective velocity linking lengths to times would have to be very small as might occur for the near localized excitations of non-periodic glassy spins.  From the temperature dependence of its dynamics, YBCO$_{6.35}$ cannot be interpreted as being close to a quantum critical point.

The saturation of the spatial and dynamic scales relates to a recent theoretical t-J model~\cite{Prelovsek05:72} that predicts that $\Gamma$ should decrease to zero upon entering the insulating antiferromagnetic phase.  Whether $\Gamma$ drops to zero continuously (in a second order manner) or discontinuously (if the quantum transition is first-order) remains to be investigated at lower doping.

Perhaps what is most surprising in Fig. \ref{mod_lor_2}  is that no clear effect of the superconductivity is observed in the excitation spectrum of YBCO$_{6.35}$  on passing through the superconducting transition.  Neither does the elastic scattering change near T$_{c}$, as noted in Section III B.  Both the inelastic and elastic response emerge continuously as the spins slow down with cooling.  In contrast, in Ortho-II YBCO$_{6.5}$ a sharp and distinct signature of coherent superconducting pairing was observed causing a suppression of the low-energy incommensurate intensity below $\hbar \omega$ $\sim$ 16 meV.  Simply scaling this superconducting gap by T$_{c}$, we would have expected to see a suppression of the magnetic scattering below energy transfers of $\sim$ 5 meV, clearly not the case in YBCO$_{6.35}$ based on our thermal and also cold neutron work.~\cite{Stock06:73}  As pointed out later with regards to the total moment sum rule, the absence of spin suppression cannot be attributed to macroscopic phase separation for we observe the expected total amount of magnetic scattering.

Our results may be associated with a microscopic decoupling of magnetic and superconducting order parameters.  The a phase transition to a state where local magnetic fields coexists  with superconductivity has been inferred by Miller \textit{et al.} using $\mu$SR by implanting the positively charged muon.~\cite{Miller06:73} In the presence of the large carrier depression caused by the strong Coulomb screening on a scale of $\sim$1 eV, in what is a very low density system near a transition to localization, we find it likely that the field, and indeed the phase, seen by or caused by the muon may differ from that detected by more gentle non-Coulomb probes such as neutron scattering and NMR.

One possibility is that locally antiferromagnetic regions are formed (within our measured correlation range) separated by metallic regions.  Although different topologically, this has the same underlying antiphase-domain physics that causes the well defined stripes observed in the nickelates.~\cite{Woo05:72}  A similar island structure has been postulated for LSCO in the insulating spin-glass region.  It could be the case that at higher hole dopings, a one-dimensional stripe structure is favored over islands which exist near the critical doping.  A related model of stripes and superconductivity addresses the different kinds of hole states.~\cite{Fine04:70}

Nonetheless our observation of a continuous evolution to a subcritical 3D spin pattern, without a transition in temperature, is at odds with a picture of physically separated islands, or of a cluster spin glass, and certainly with phase separation, for which the highly organized 3D spin pattern along [0 0 1] would not occur.  In particular the continuous transfer of spectral weight from the relaxational to the central mode, as discussed later (Fig. \ref{integral}), is a strong argument in favor of a single phase in which slow short-range glassy order and spin excitations coexist with superconductivity.  The absence of a spin anomaly at  T$_{c}$ at low doping indicates that it is the holes rather than the coherence of pairs that break up the underlying antiferromagnetic spin pattern.  Only at larger doping T$_{c}$ = 59 K with stronger superconducting order, is the onset of pair coherence large enough to cause a superconducting gap in the spin response.

\subsection{$\omega/T$ Scaling}

\begin{figure}[t]
\includegraphics[width=93mm]{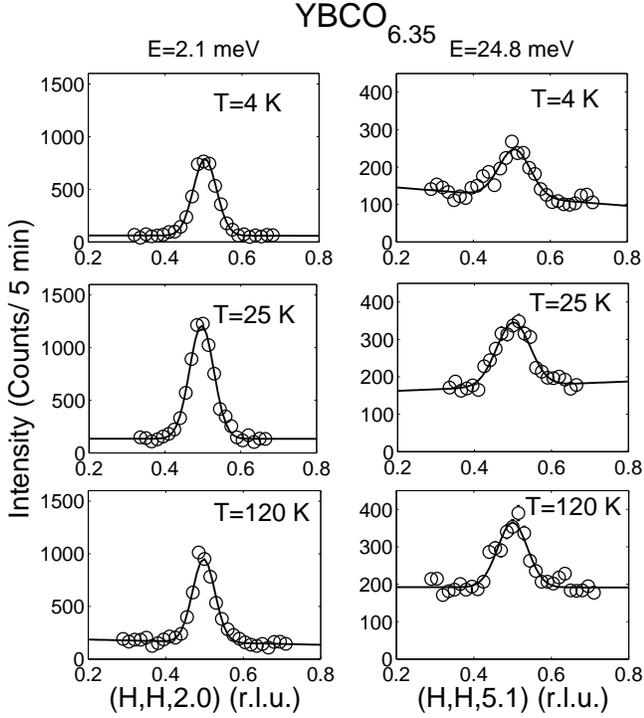}
\caption{Constant energy scans for $\hbar \omega$=2.1 meV and 24.8 meV for a series of temperatures.  The inelastic scattering is commensurate with no observable change in width as a function of temperature and energy transfer.  The solid lines are the results of fits to Gaussians with a sloping background.  The $\omega/T$ scaling analysis was conducted based on Gaussian fits such as those represented here.} \label{scaling_2}
\end{figure}

\begin{figure}[t]
\includegraphics[width=85mm]{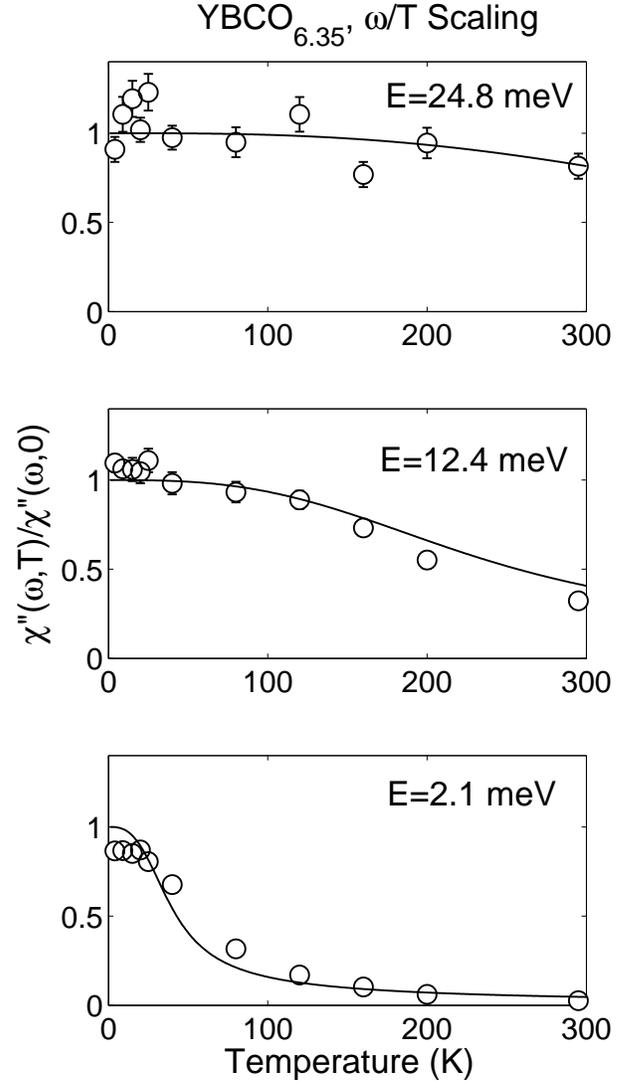}
\caption{The normalized susceptibility $\chi''(\omega,T)/\chi''(\omega,0)$ is plotted as a function of temperature at 2.1, 12.4, and 24.8 meV.  The solid lines are results of fits to the $\omega/T$ scaling analysis described in the text.  The fit describes the data over a broad range in temperature and energy.} \label{scaling_3}
\end{figure}

\begin{figure}[t]
\includegraphics[width=90mm]{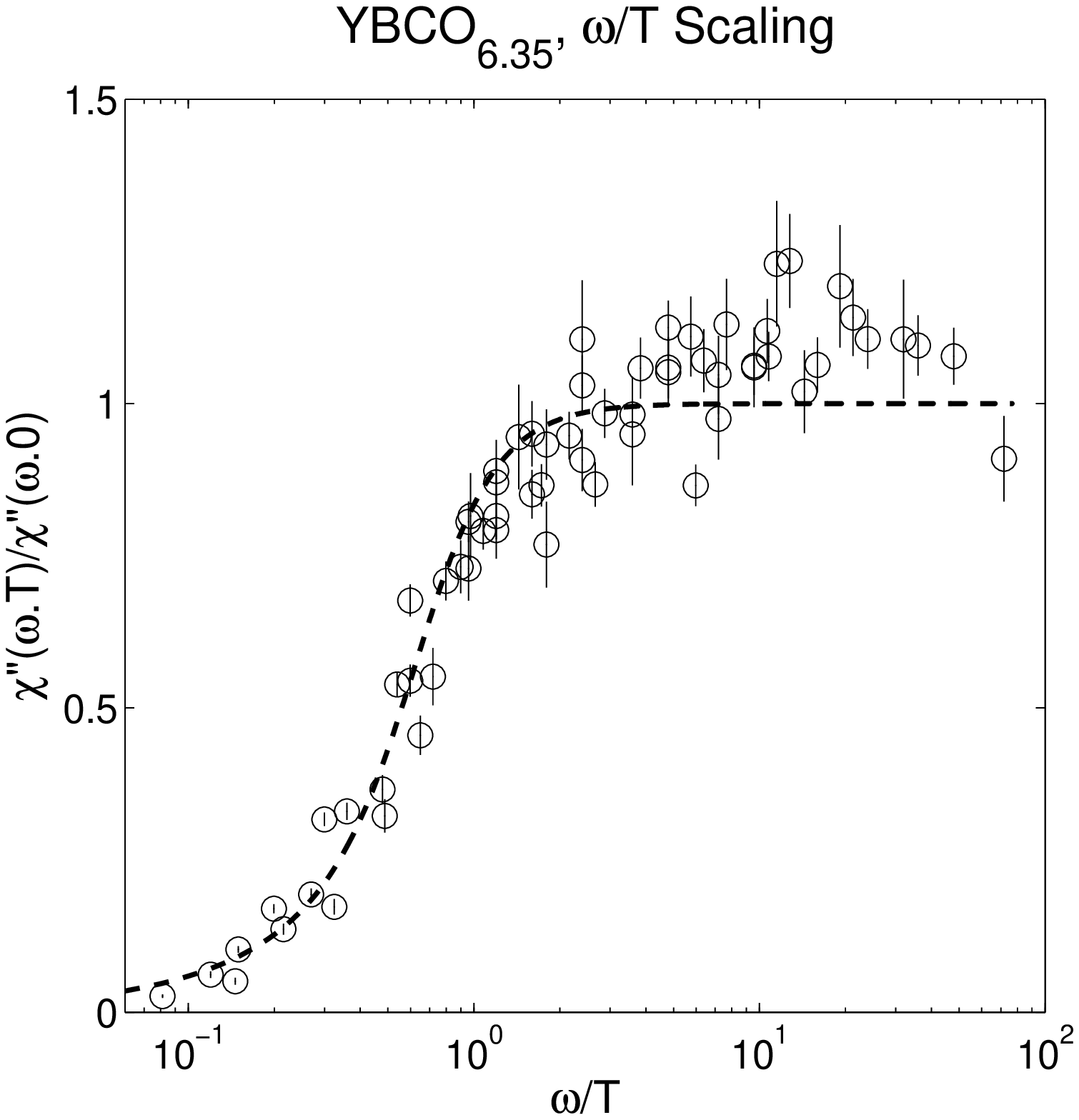}
\caption{A plot of the normalized $\chi$$'$$'$ derived from constant energy scans for all energy transfers studied ($\hbar \omega$ = 2.1, 3.7, 8.3, 12.4, 16.5, and 24.8 meV).  The solid lines are results of fits to the scaling equation discussed in the text.  The $x$-axis is plotted on a logarithmic scale.} \label{scaling_1}
\end{figure}

    In early studies $\chi$$'$$'$ in disordered YBCO$_{6.5}$ was found to follow a simple $\omega$/T scaling relation over a broad range in energy transfers and temperatures.~\cite{Birgeneau92:87}  The same scaling relation successfully describes the temperature dependence of the local susceptibility in underdoped La$_{2-x}$Sr$_{x}$CuO$_{4}$.~\cite{Keimer91:67,Hiraka01:70}  The scaling relation, which assumes that the dominant energy scale is set by the temperature, predicts the following temperature and energy dependence for the susceptibility,

\begin{eqnarray}
{\chi''(\omega, T) \over \chi''(\omega,T=0)}= {2 \over \pi}\arctan \left({a_{1} \omega \over T}+{a_{2} \omega^{3} \over T^{3}}+...\right).
\label{scaling_all}
\end{eqnarray}

\noindent Here the term $\chi$$'$$'$$(\omega,T=0)$ represents the limit of $\chi$$'$$'$ as the temperature goes to zero.  In Ortho-II ordered YBCO$_{6.5}$  the temperature dependence at all energies up to $\sim$ 20 meV is well described by only the first term in the expansion with coefficient ($a_{1}$).  A similar expansion describes the the excitations up to $\sim$ 30 meV in disordered YBCO$_{6.5}$.~\cite{Birgeneau92:87}  Authors of recent studies of hole doped and electron doped cuprates have chosen to use another form of scaling analysis where $\chi''(Q_{AF},\omega,T)$ is plotted against $\omega/T$ to find a single universal curve.~\cite{Bao03:91,Wilson06:74}  This particular analysis is a low-energy approximation and has been presented in connection with the possibility of the close proximity of a quantum critical point.  We have chosen to use the well established analysis which assumes only the less restrictive hypothesis that temperature is the dominant energy scale and so should apply to all energy transfers.

To obtain an amplitude (and hence $\chi$$'$$'$) as a function of energy and temperature, constant energy scans were conducted along the [110] direction for a fixed temperature and energy transfer.  The value of L was chosen based on the bilayer structure factor to maximize the scattered intensity.  For our scaling analysis, temperatures in the range of 2 - 293 K were investigated and energy transfers in the range of 2.1 - 24.8 meV measured.  Elastic scattering does not figure in the scaling analysis. Typical scans over a broad range in temperature and energy are presented in Fig. \ref{scaling_2}.  The correlated scattering is well described by the fitted Gaussians.  We have also made use of the fact that the momentum width is found to be independent of energy and temperature.   Each point in the scaling analysis is then based on a single amplitude parameter from the constant energy scan and is proportional to the local susceptibility.  Recent results  show that the momentum profile broadens and evolves at high energies into two peaks whose velocity is the same as that observed in the cuprate insulator~\cite{Stock07:75} but this occurs outside the range $\hbar \omega <$ 35 meV of the current experiment.

The normalized susceptibility  (defined as $\chi''(\omega,T) / \chi''(\omega,0)$) is plotted in Fig. \ref{scaling_3} for three typical energy transfers.  We then simultaneously fit the scaling equation of Eqn. \ref{scaling_all} to all data   as shown by the broken line in Fig. \ref{scaling_1}.  The entire susceptibility was best described by the inclusion of two terms in the scaling analysis:

\begin{eqnarray}
{\chi''(\omega, T) \over \chi''(\omega,T=0)}= \nonumber \\
{2 \over \pi} \arctan \left({0.90\pm 0.04\left(\omega \over T\right)}+{2.8\pm0.3 \left(\omega \over T \right)^3}\right).
\end{eqnarray}

\noindent The inclusion of higher terms in the expansion did not noticeably improve the fits.  The solid lines in Fig. \ref{scaling_3} are typical predictions of the scaling equation.    It is seen that $\omega/T$ scaling gives an excellent description of the data over several decades.  The value of $a_{1}$ is very close to the value of 1.1 obtained in the YBCO$_{6.5}$ Ortho-II superconductor for temperatures above T$_{c}$ and energies well below the resonance energy.  The values for a$_{1}$ and a$_{2}$ differ considerably from those obtained in LSCO for similar hole doping.~\cite{Keimer92:46}  While this implies a common temperature dependence to the magnetic fluctuations over a wide range in the underdoped YBCO$_{6+x}$ phase diagram, the scaling parameters do not necessarily universally apply to other cuprate superconductors.

Even though scaling is approximately obeyed over a broad energy and temperature range, the lowest panel of Fig. \ref{scaling_3} illustrates that the susceptibility at 2.1 meV does show evidence for a subtle breakdown below 50 K.  We attribute this breakdown to the low-temperature region where the energy full-width $2\Gamma$ ceases to decline linearly with temperature.  The reduction below scaling in the susceptibility at 2.1 meV below 50 K occurs at the temperature where the weight in the relaxational spin excitation begins to transfer to the quasielastic central mode as will be seen (Fig. \ref{integral}).

Previously in lightly doped insulating LSCO, a similar breakdown of $\omega/T$ scaling was observed.  The energy scale required to account for the breakdown was associated with an out-of-plane anisotropy which causes a gap in the excitation spectrum and hence a new energy scale.~\cite{Matsuda93:62}  In YBCO$_{6.35}$ we cannot attribute the breakdown of scaling to the presence of such an anisotropy as we observe the elastic correlations to be paramagnetic or isotropic, in contrast to the behavior of the insulator.  However, the new energy scale required to account for the breakdown of scaling can be attributed to the confinement of the spins as discussed in the context of correlated regions in the previous section.

In Ortho-II YBCO$_{6.5}$ scaling is obeyed in the normal state above 59 K and for energy transfers below $\sim$ 20 meV.  The breakdown of scaling at high-energy transfers in ortho-II was attributed to the presence of a resonance peak at 33 meV introducing a new energy scale.  The breakdown for temperatures below the onset to superconductivity was characteristic of the superconducting gap opening and suppressing magnetic fluctuations below 16 meV.  Neither a superconducting gap, nor a strong resonance peak is observed in YBCO$_{6.35}$, and therefore scaling is observed over a broad range in temperature and energy.  Theoretical motivation for scaling can be found in theories of marginal Fermi liquids (Ref. \onlinecite{Varma89:63}), close proximity of a quantum critical point (Ref. \onlinecite{Bao03:91}), stripe liquid (Ref. \onlinecite{Zaanen97:282}), damping of a collective mode (Ref. \onlinecite{Prelovsek04:92}), and from a memory-function spin-wave theory(Ref. \onlinecite{Larionov05:72}).

The fact that we observe scaling over a broad energy and temperature range in YBCO$_{6.35}$ shows that any substantial spin pseudogap is absent.  In contrast a charge pseudogap has been investigated  for more highly doped YBCO$_{6+x}$ by optical and NMR measurements which reveal a suppression of 1/T$_{1}T$ and of the optical conductivity below 30 meV.~\cite{Hwang06:73,Timusk99:62}  The presence of a spin pseudogap would imply that spin response $\chi''$ would be suppressed at low temperatures if it were to follow the NMR relaxation 1/T$_{1}T$ or the optical conductivity.  On the contrary the staggered spin susceptibility and instead grows monotonically on cooling and we observe scaling and moment conservation over a broad range in temperatures.  Sutherland \textit{et al.} recently suggested from transport measurements that the pseudogap energy scale may lie at $\sim$ 160 meV in low-doped materials.~\cite{Sutherland05:94}  Therefore, the effects of the pseudogap are likely only felt at very high energies as recently revealed by the large spin spectral weight loss above 160 meV in the high-energy magnetic excitations.~\cite{Stock07:75}

\section{Integrated Intensity and Sum Rule}

\begin{figure}[t]
\includegraphics[width=80mm]{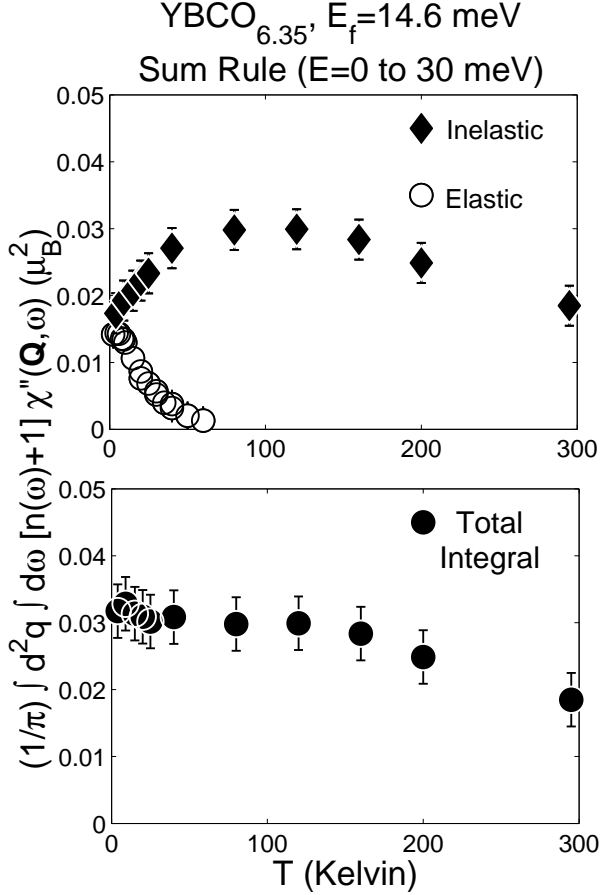}
\caption{The integrated intensity in absolute units as a function of temperature for energy transfers below 30 meV.  The upper panel shows the transfer on cooling of weight from the broad inelastic relaxational Lorentzian to the central peak around $\hbar \omega$=0.  The lower panel depicts the total integral and shows that the sum rule is conserved at low temperatures.  The decrease in total spectral weight at high temperatures is due to the broad energy width of the modified Lorentzian.} \label{integral}
\end{figure}

To obtain a quantitative measure of the spectral weight, and of the fraction of the total moment which resides in the low-energy central peak, we consider the sum rule for two \textit{uncoupled} CuO$_{2}$ layers.  Ignoring the effect of the chains, and therefore taking only two Cu$^{2+}$ ions per formula unit, the total spectral weight integrated over all energy and momentum is given by the total moment sum rule:

\begin{eqnarray}
\label{sum_rule_S} \int d\omega\int d^{3}q \ {S(\bf{q}, \omega)}=2\times {2 \over 3} S (S+1) g^{2} ,
\end{eqnarray}

\noindent  where $\int d^{3}q$ is the momentum space integral over the correlated peak.  In this equation the factor of 2 comes from the fact that we are taking two Cu$^{2+}$ ions per formula unit.  Here we have assumed that $S(\bf{q}, \omega)$ has been corrected for the bilayer structure factor and the Cu$^{2+}$ anisotropic form factor.  Doubtless the total moment sum rule will be somewhat reduced as some of the spins will be destroyed by the doped holes.  Since the doping is so small, $p$ $\sim$ 0.06 for YBCO$_{6.35}$, and since the oxygen holes that screen the Cu$^{2+}$ are not seen at neutron momenta, we expect the total moment sum rule to be approximately obeyed.  Equating the above integral to the cross-section for paramagnetic scattering

\begin{eqnarray}
\label{sum_rule_chi}
I \equiv \pi^{-1} \int d\omega \int d^{3}q  [n(\omega)+1]{\chi''(\bf{q}, \omega)}= \\
\nonumber {2 \over 3} \mu_{B}^{2} g^{2} S (S+1).
\end{eqnarray}

\noindent gives for S=$1 \over 2$ and $g=2$ a total integral of I$=$ $2 \mu_{B}^{2}$.  This limit provides a useful benchmark with which we can compare our integrated intensities. We determined the absolute intensity with two methods.  All elastic and inelastic intensities were calibrated to the known constant-Q integral over a transverse acoustic phonon at (0.15, 0.15, 6) as described in Ref. \onlinecite{Stock04:69}.  This gives one value for the calibration constant.  As an independent check, we also calibrated just the elastic scattering by measuring a series of Bragg intensities corrected for extinction, and fitting the calibration constant as described in the Appendix.  Both methods agree well for the elastic scattering justifying the quantitative analysis used here.

In YBCO$_{6.35}$, at 2 K, we observe a total moment below 30 meV of 0.031 $\pm$ 0.006 $\mu_{B}^{2}$.  The apportionment of the integrated spectral weight to the central peak and to the modified Lorentzian is given in the upper panel of Fig. \ref{integral}.  The lower panel shows that the total integral (sum of central peak and modified Lorentzian) is conserved at all low temperatures below 150 K thus confirming the total moment sum rule (response above 30 meV$\sim{350K}$  is not expected to change at these temperatures).  The small reduction at high temperatures is a result of the energy scale of the modified Lorentzian becoming larger than the integration range of 0 to 30 meV; it does not represent a loss of integrated spectral weight.

As the temperature falls below $\sim 50$ K Fig. \ref{integral} shows that spectral weight transfers from the broad inelastic scattering to the central peak  its gain being fully compensated by the loss of weight from the excitations.  The conservation of spin weight is arguably the clearest evidence that we are dealing with a single phase system. The slowing of the relaxational response couples to and drives up the central peak as demonstrated by the excellent description provided by our coupled soft mode central mode model.~\cite{Stock06:73}.  The model corresponds to inelastic excitations and a central peak that arise from the same single bulk phase. The extremely slow dynamics of the central mode can be viewed as the slow tumbling of the many correlated spins, and the fast inelastic relaxation as the excitations of these glassy spins. The central peak cannot be attributed to an impurity or macroscopic phase separation.   Macroscopic phase separation has been cited as a possible reason for the coexistence of static magnetic order and superconductivity in La$_{2}$CuO$_{4+y}$~\cite{Hammel90:42}, a view not supported by our data on high-quality crystals.

    The transfer of spectral weight from the inelastic channel to the central peak occurs at the same temperature as the saturation in the dynamic linewidth where we observe a breakdown of $\omega/T$ scaling for low energies.  The breakdown arises from the transfer to the gradually slowing down lowest-energy spins, forming a new energy scale represented by the central peak and not included in the scaling analysis.  This conclusion differs from the scaling analysis performed on Li doped La$_{2}$CuO$_{4}$ which attributes the breakdown of $\omega/T$ scaling to the close proximity of a quantum critical point.~\cite{Bao03:91}  Nonetheless there is a critical point to antiferromagnetism in the vicinity and, despite very short range correlations, the central mode we see foretells its arrival at lower doping.

\begin{figure}[t]
\includegraphics[width=80mm]{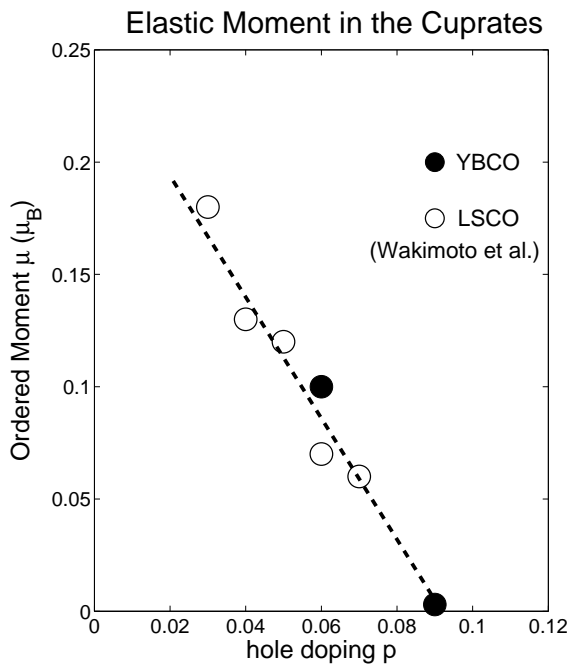}
\caption{The low temperature magnetic moment observed with elastic thermal neutron scattering as a function of the hole doping for both YBCO and LSCO~\cite{Wakimoto01:63} systems. The YBCO point at $p$=0.09 is an upper bound as described in the text.~\cite{Stock02:66}  The uniform suppression of the elastic magnetic scattering with no anomaly on entry to the superconducting state at $p_{c}$=0.055 indicates that holes not pairs destroy the moment.} \label{hole_doping}
\end{figure}

    In the YBCO$_{6+x}$ system, the in-plane exchange constant is estimated to be J $\sim$ 125 meV implying that the spin excitations would extend to about $\hbar \omega$ $\sim$ 250 meV.  If the spectral weight were evenly distributed over all energies we would expect to observe (30 meV/250 meV) $\times$ 2 $\sim$ 0.24 $\mu_{B}^{2}$ which is far above what is seen experimentally in YBCO$_{6.35}$.  Even recognizing that the density of states is weighted to high energies in a 2D antiferromagnet, we calculate that 0.06 of the spin spectral weight lies below 30 meV, a calculation known to be too large because of the errors in spin-wave theory.  We would then expect to observe 0.06 $\times$ 2 $\mu_{B}^{2}$ $\sim$ 0.12 $\mu_{B}^{2}$, still much larger than the 0.03 we observe.  To obtain a better estimate of the expected total spectral weight, we rely on the spectral density of 4 $\mu_{B}^{2}/eV$ obtained in insulating YBCO$_{6.15}$.~\cite{Hayden96:54}  Taking the low-energy limit where the spin excitations disperse linearly with energy, the total expected integrated moment should be $\sim$ 4 $\mu_{B}^{2}/eV$ $\times$ 30 meV /$\pi$ $\sim$ 0.04 $\mu_{B}^{2}$.  This agrees within error with the total integral of 0.03 $\mu_{B}^{2}$ measured in YBCO$_{6.35}$.  The fact that we get good agreement with the total moment of the antiferromagnetic insulator provides additional evidence that the central peak is a necessary macroscopic property of single-phase YBCO$_{6.35}$.

    The total integrated spectral weight up to 30 meV in YBCO$_{6.35}$ agrees very well with that observed in more highly doped Ortho-II YBCO$_{6.5}$ despite the fact that the latter lies further from the antiferromagnetic insulating phase.  As the integral from 0 to 40 meV in ortho-II gave 0.05 $\mu_{B}^{2}$, we would expect that the integral from 0 to 30 meV in YBCO$_{6.35}$ should give $\sim$ 0.04 $\mu_{B}^{2}$ in good agreement with what we measure.

     An important conclusion is that, within the YBCO$_{6+x}$ phase diagram, our results demonstrate that the total moment is conserved not only with temperature, but also with doping from the insulating region to deep within the superconducting phase.  Therefore, the spectral weight is not removed on entry into the superconducting phase, but simply redistributed in energy and momentum.  From YBCO$_{6.35}$ to YBCO$_{6.5}$, more spectral weight is moved from the elastic and $\sim$  2 meV region to higher energies around the resonance energy at 33 meV (as illustrated in Fig. \ref{figure1}).  The apparent lack of spectral weight at larger doping may be associated with lack of visibility caused by a broadening in momentum and not necessarily by a removal of spectral weight.~\cite{Wakimoto04:92}  A detailed analysis of the total moment and various contributions to the integral has been conducted by Lorenzana \textit{et al.}~\cite{Lorenzana05:72}

    In Fig. \ref{hole_doping} we plot the dependence of the low temperature statically ordered moment on hole doping for both YBCO (see Ref.~\onlinecite{Stock05:71} and this work) and LSCO~\cite{Wakimoto01:63}. The moment includes the central peak, previously shown to be quasielastic~\cite{Stock06:73}, and integrated over its momentum breadth.  The ordered moment was obtained by calibrating against Bragg peaks as described in the Appendix.   For YBCO$_{6.35}$ and YBCO$_{6.5}$ the ordered moment was obtained from a comparison with the nuclear Bragg peaks.   The datum for YBCO$_{6.5}$, where no magnetic Bragg peak is observable~\cite{Stock02:66}, is an upper bound placed using neutrons and has been confirmed by NMR.~\cite{Yamani04:405}  With increase of doping there is no break in the uniform suppression nor an anomaly as the superconducting phase is entered.  We infer that it is the accommodation of holes by spin reorganization rather than pairing of charges that destroys the static moment.

Significantly Fig. \ref{hole_doping} shows that the $q$-integrated quasielastic moment in YBCO$_{6.35}$ maps directly onto the curve for the measured incommensurate peaks in LSCO.  It is surprising then that transport measurements at milli-Kelvin temperatures~\cite{Sutherland05:94} imply that LSCO and YBCO have very different carrier densities for similar hole doping (LSCO is more localized).  In contrast the static ordered moment on the THz timescale) indicates that LSCO and YBCO are very similar.  This conclusion needs to be verified in YBCO to determine what differences may exist between the LSCO and YBCO phase diagrams nearer the critical hole doping of $p_{c}$=0.055.

\section{Discussion}

    We have shown that the spin dynamics are governed by two distinct energy scales, one described by an energy resolution limited central peak indicating slow fluctuations which become static at the lowest temperature on the thermal neutron time scale, and a second fast component with a low temperature characteristic energy of $2\Gamma$=5 meV.  The width of the fast component scales with temperature above 50 K and saturates at low temperatures.  We observe no well-defined resonant feature as occurs for larger hole doping.  The momentum dependence illustrates that the correlations of the fast inelastic component are highly 2D in nature.

The slow fluctuations, in contrast, are not described by a single relaxation rate and display 3D correlations at low temperature.  The correlation lengths (determined from a Lorentzian squared) are short-ranged with $\xi_{ab}$= 13 $\pm$ 2 \AA\ and $\xi_{c}$=8 $\pm$ 2 \AA\. The spins exhibit isotropic orientation. The quasielastic scattering  corresponds not to static N\'{e}el order, but to a gradual slowing down of the dynamics with temperature.  The slow component only becomes appreciable at low temperatures below $k_{B}T$ $\sim$ 50 K $\sim$ 5 meV, where the dynamic width of the fast component saturates.  It is not clear what sets the energy scale of the fast component but we regard it as the excitation spectrum of a short-range spin glass structure.

    The results cannot be attributed to the presence of macroscopic phase separation of a superconducting region and antiferromagnetic regions with a range of hole doping.   Our samples show a single sharp superconducting transition which rules out the presence of other superconducting domains with different hole doping.  If phases with a range of N\'{e}el temperatures were present we would have observed resolution limited peaks at the antiferromagnetic wave vector. Instead we observe an elastic cross section that is broad in momentum along the c-axis and also within the $a-b$ plane.  The most telling evidence for single-phase behavior is that the central peak gathers spectral weight as the inelastic channel loses weight. The two features are therefore dynamically coupled while conserving total moment confirming an earlier hypothesis.~\cite{Stock06:73} This conclusion is further verified by the single lattice constant measured by X-rays in our samples.  As shown by Liang \textit{et al.}, the lattice constants (and in particular the $c$-axis) are very sensitive to hole doping in the heavily underdoped region of the phase diagram.~\cite{Liang06:73}  We do not observe a spread in lattice constants and therefore conclude that our sample is homogeneous.

   The work presented here is very different from early studies conducted in the underdoped region of the phase diagram.  Work by J. Rossat-Mignod \textit{et al.} did not investigate the region of the phase diagram with a T$_{c}$ less than 50 K and did not observe directly the two dynamically coupled timescales differentiated in the present measurements.~\cite{Mignod93:192} Their samples were not superconducting for oxygen content less than 6.40. More recent work by Mook \textit{et al.} on a oxygen concentration of $x$=0.35 was performed on a sample with a much higher T$_{c}$ than discussed here, implying a larger hole doping in the CuO$_{2}$ planes.~\cite{Mook02:88}  It follows that the samples studied here have a hole concentration much closer to the critical value of $p_{c}$=0.055 than previous studies.

    The  quasielastic scattering is similar to that in the monolayer La$_{2-x}$Sr$_{x}$CuO$_{4}$ near the critical doping of $p_{c}$=0.055 and so illustrates that the properties are not unique to the monolayer cuprates.   One difference is that we observe commensurate magnetic peaks whereas the monolayer cuprates display incommensurate peaks.  Since the resolution was adequate to detect a similar elastic incommensurate wave vector, it may be that the incommensurability in underdoped YBCO$_{6.35}$ is smaller or suffers intrinsic broadening because of short oxygen chain order.  Nonetheless we have shown that the elastic correlation lengths and quasielastic moment are similar to the monolayer  cuprate system.   The total spectral weight at energies below 30 meV energy transfer has a similar magnitude to that in the insulating cuprates.    We find that as the critical hole concentration is approached from above, slowly dynamic antiferromagnetic short-range order develops, as in a glass, at the expense of superconductivity.

     A spectrum with two energy scales similar to that measured here in hole-doped YBCO$_{6.35}$ also occurs in electron-doped  Pr$_{0.88}$LaCe$_{0.12}$CuO$_{4-\delta}$ (PLCCO) with T$_{c}$=21 K.~\cite{Wilson06:74,Wilson06:96}    This adds credence to our suggestion that the two energy scales with central mode as observed in the YBCO$_{6+x}$, LSCO, and PLCCO systems are a generic feature and common to underdoped high-temperature superconductors.

    The low-energy excitation spectrum in YBCO$_{6.35}$ is very different from that of more heavily doped samples.  The low-energy inelastic scattering is commensurate with no sign of a well-defined resonance peak.   In the Ortho-II YBCO$_{6.5}$ superconductor the incommensurate modulation is dynamic within the hour-glass.~\cite{Stock04:69} A suppression of magnetic scattering below $\sim$ 16 meV below T$_{c}$, indicates the formation of a superconducting spin gap. No such gap or suppression is observed in YBCO$_{6.35}$, but instead an inexorable growth through T$_{c}$.

        Based on previous doping studies, we conclude that the lack of a clear resonance and spin-gap \textit{cannot} be attributed to the presence of structural disorder.  It may be possible that disorder present in the chains could localize charge and have an effect on transport properties in the CuO$_{2}$ planes.~\cite{Franco03:67}  Effects of structural disorder on the superconductivity have been noted in excess oxygen La$_{2}$CuO$_{4+y}$ by Lee \textit{et al.}~\cite{Lee04:69}  Doping impurities into the YBCO$_{6+x}$ superconductor has also been shown to induce long-ranged magnetic order.  For cobalt doping, the cobalt ions enter the chains introducing not only structural disorder but also antiferromagnetic order, even in heavily hole doped samples.~\cite{Hodges02:66}  The order was long-ranged characterized by sharp Bragg peaks when scanned along both the [H,H,0] and [0,0,L] directions. Long-range order is patently not present in our YBCO$_{6.35}$ samples which display short correlation lengths both along $c$ and within the $a-b$ planes and a highly organized 3D spin pattern that indicates that significant structural disorder is absent.   While impurity and disorder effects have been characterized in heavily doped YBCO to have subtle effects on the lineshape and gap structure, they do not dramatically alter the lineshape and certainly do not remove the resonance and the spin-gap.  Therefore, disorder cannot account for the large differences we observe in YBCO$_{6.35}$ and more heavily doped YBCO$_{6.5}$.


    While the doped holes do not have a strong effect on integrated strength of low-energy excitations, they nearly destroy the high-energy excitations above $\sim$ 200 meV as they enter the pseudogap near the zone boundary.~\cite{Stock07:75}  The low-energy fluctuations (characteristic of long-wavelength fluctuations) and the central peak are therefore sensitive to spin disorder introduced through the doping of holes.  It is only at high energies where the decay of of spin excitations into particle hole pairs is energetically possible, that the spin excitations become sensitive to electronic degrees of freedom.  We note that the spin properties measured here are very different from those of heavily doped samples.  This observation may be in accord with thermal conductivity experiments which have found a metal insulator transition in the YBCO$_{6+x}$ system for $x$ near 0.5.~\cite{Sun04:93}  Indeed, a detailed study of the superconducting spin gap as a function of hole doping in YBCO suggests that the superconducting gap decreases rapidly for small hole doping.~\cite{Dai01:63}

\section{Conclusions}

We have presented a detailed study of the static and dynamic magnetic properties of YBCO$_{6.35}$ (T$_{c}$=18 K).  Two distinct timescales exist, one a fast relaxational excitation with an energy linewidth of $2\Gamma$=5 meV, and a second slowly dynamic centra mode that appears static at low temperatures on the neutron timescale.  In contrast to what is observed in more heavily doped samples, we do not observe a clear signature of superconductivity in the temperature dependence of the dynamic susceptibility, indicating  a decoupling of the magnetic Cu$^{2+}$ spins from the paired quasiparticles but not from the holes.  For temperatures greater than 50 K, the magnetic scattering is well described by $\omega/T$ scaling with the spectral energy width scaling linearly with temperature.  A breakdown of scaling is observed at low temperatures and energies where the central peak develops and the spins gradually freeze into a spin glass state removing spectral weight from the excitations.  We find  overall that the behavior of a cuprate superconductor close to its critical doping is dramatically different from that of the heavily doped superconductors and involves a large shift of the spectral weight to low and quasielastic energies.

\section{Acknowledgements}

We thank R.A. Cowley, J.M. Tranquada, S.-H. Lee and C. Broholm for helpful comments and discussions, and M. Potter, L. McEwan, R. Sammon, R. Donaberger and J. Bolduc for excellent technical support during experiments at CNBC-NRC, Chalk River.  The work at Chalk River and Johns Hopkins University was supported by the Natural Sciences and Engineering Research Council (NSERC) of Canada, a GSSSP scholarship through the National Research Council of Canada, and through DMR 0306940.  The work at Lawrence Berkeley Laboratory was supported by the Office of Basic Energy Sciences, US Department of Energy under contract number DE-AC03-76SF0098.

\section{Appendix: Absolute Calibration using Bragg peaks}

\begin{figure}[t]
\includegraphics[width=80mm]{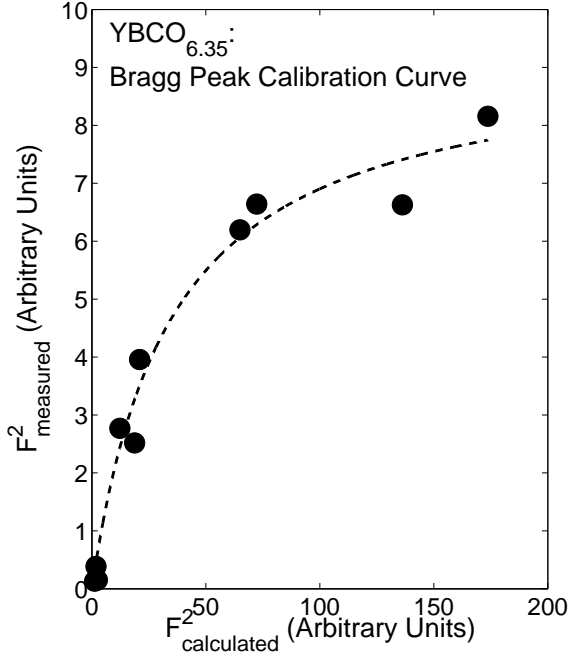}
\caption{The measured nuclear structure factor squared as a function of calculated structure factor squared is plotted for YBCO$_{6.35}$.  The data points are based on the integrated intensity calculated from transverse scans through Bragg peaks.  The curve is the result of a fit to the empirical formula in the text that accounts for secondary extinction.} \label{bragg_cal}
\end{figure}

\begin{figure}[t]
\includegraphics[width=80mm]{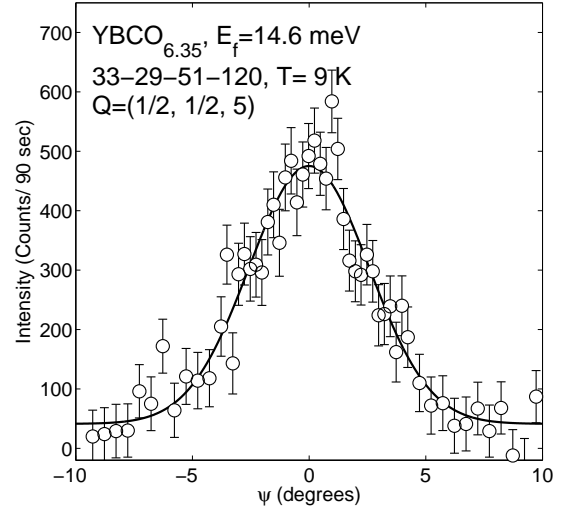}
\caption{A transverse ($\psi$) scan through the correlated magnetic scattering at (1/2, 1/2, 5). The data was obtained by subtracting a high temperature background taken at 80 K.} \label{bragg_omega}
\end{figure}

To obtain an estimate of the strength of the quasielastic magnetic moment in YBCO$_{6.35}$ we carried out, in addition to the phonon calibration,  an absolute calibration against the nuclear Bragg peaks.  This estimate allows a direct comparison with ordered antiferromagnets to determine how much of the Cu$^{2+}$ moment is ordered quasi-statically.  Statically here means including the slow quasielastic peak since there is no strictly elastic scattering in YBCO$_{6.35}$.  The Bragg peak structure factor was determined from rocking scans in crystal angle $\psi $ using the relation

\begin{eqnarray}
\label{sum_rule_chi}
\int d\psi   \ I_{meas.}(\psi) = {|F|^{2} \over {\sin(2\theta)}}.
\end{eqnarray}

\noindent To carry out the calibration we plot the calculated structure factor of the nuclear Bragg peaks against the measured structure factors to obtain a calibration constant.  The resulting curve is plotted in Fig. \ref{bragg_cal} and results from a fit to the following equation to correct for secondary extinction

\begin{eqnarray}
\label{sum_rule_chi}
|F_{measured}|^{2} = {{\alpha |F_{calculated}|^{2}} \over{1+\beta |F_{calculated}|^{2}}}.
\label{equation_bragg}
\end{eqnarray}

\noindent The fitted calibration constant is $\alpha$  and $\beta$ denotes a factor to account for extinction.  In the limit of small $\beta$, Eqn. \ref{equation_bragg} agrees with the expression of Shamoto \textit{et al.}~\cite{Shamoto93:48}  Given the complicated geometry of our sample mount consisting of seven co-aligned crystals, this simple model fits remarkably well.

The structure factor for the integrated intensity for two copper spins per unit cell is given by

\begin{eqnarray}
\label{structure_factor}
{|F_{M}|^{2}={2\over 3}{(\gamma r_{0})^{2}\over 4} \mu^{2} g^{2}f^{2}({\bf{Q}})B^{2}({\bf{Q}})}.
\end{eqnarray}

\noindent where $(\gamma r_{0})^{2}/4$ is equal to 73 mbarns, $g$ is the Lande factor set to 2, $B^{2} ({\bf{Q}})$ is the bilayer structure factor previously defined, and $\mu$ is the value of the ordered moment.  The factor of $2/3$ comes from the polarization factor for isotropic spins as found from our polarized neutron analysis.  The integral of the rocking scan of the central peak was obtained by subtracting from a 9 K scan a non-magnetic background at 80 K.  This is the same subtraction method used throughout this study to obtain the elastic correlated magnetic scattering.  By comparing the rocking scan through the (1/2, 1/2, 5) (shown in Fig. \ref{bragg_omega}) position with the Bragg peak intensity, we obtain an ordered square moment of  $\mu^{2}$=0.010$\pm$0.002 $\mu_{B}^{2}$ in YBCO$_{6.35}$ compared with $\sim$ 0.47 $\mu_{B}^{2}$ measured by Shamoto \textit{et al.} in insulating YBCO$_{6.15}$.  Therefore, the square of the static moment has decreased by an order of magnitude from insulating YBCO$_{6.15}$ to superconducting YBCO$_{6.35}$.  A similar decrease in the ordered moment has been observed in the single layer LSCO system as a function of hole doping.~\cite{Wakimoto01:63}

    The method used here to calibrate the elastic scattering assume the peak is resolution limited along the longitudinal direction.  This is clearly not the case and may account for the slight underestimate of the elastic moment when compared with the calibration analysis based on an acoustic phonon.  Nevertheless, both methods agree well and provide justification for the quantitative analysis presented here.  The value for the total ordered moment also provides a useful comparison with the insulator and other cuprate systems.

\end{document}